\documentclass{hep99}
\usepackage{graphicx}
\usepackage{epsfig}
\def    \nn             {\nonumber}
\def    \=              {\;=\;}
\def    \frac           #1#2{{#1 \over #2}}

\def    \lsim           {\raisebox{-3pt}{$\>\stackrel{<}{\scriptstyle\sim}\>$}}
\def    \gsim           {\raisebox{-3pt}{$\>\stackrel{>}{\scriptstyle\sim}\>$}}

\newcommand     \be     {\begin{equation}}
\newcommand     \ee     {\end{equation}}
\newcommand     \ba     {\begin{eqnarray}}
\newcommand     \ea     {\end{eqnarray}}
\newcommand     \sst            {\scriptstyle}   
\newcommand     \sss            {\scriptscriptstyle}

\newcommand     \epem           {\ifmmode{e^+e^-}\else{$e^+e^-$}\fi}
\newcommand     \lambdamsb     {\ifmmode
          \Lambda_4^{\rm \scriptscriptstyle \overline{MS}} \else
         $\Lambda_4^{\rm \scriptscriptstyle \overline{MS}}$ \fi}
\newcommand     \MSB            {\ifmmode {\overline{\rm MS}} \else 
                                 $\overline{\rm MS}$  \fi}

\newcommand     \ptmin     {\ifmmode p_{\scriptscriptstyle T}^{\sss min} \else
                           $p_{\scriptscriptstyle T}^{\sss min}$ \fi}

\newcommand\as{\alpha_{\sss S}}         
\newcommand\astwo{\alpha_{\sss S}^2}         
\newcommand\asthree{\alpha_{\sss S}^3}         
\newcommand\asfour{\alpha_{\sss S}^4}

\newcommand\epspoeps{\ifmmode \frac{\epsilon'}{\epsilon}
              \else   $\epsilon'/\epsilon$ \fi} 

\def \oatwo {\mbox{${\cal O}(\astwo)$}}

\def \pt   {\mbox{$p_{\scriptscriptstyle T}$}}                
\def \et   {\mbox{$E_{\scriptscriptstyle T}$}}

\def \to   {\mbox{$\rightarrow$}}

\newcommand \vckm{\ifmmode{V_{\rm CKM}
    }\else{$V_{\rm CKM}$}\fi} 
\newcommand \jpsi{\ifmmode{J/\psi
    }\else{$J/\psi$}\fi} 
\newcommand \cpviol{\ifmmode{\not
    \!\!\!{\rm CP}}\else{$\not \!\!\!{\rm CP}$}\fi}

\def\calF{{\cal F}}
\def\calP{{\cal P}}
\def\calM{{\cal M}}
\def\calO{{\cal O}}
\def\gappeq{\mathrel{\rlap {\raise.5ex\hbox{$>$}}
{\lower.5ex\hbox{$\sim$}}}}
\def\lappeq{\mathrel{\rlap{\raise.5ex\hbox{$<$}}
{\lower.5ex\hbox{$\sim$}}}}
                         
\def\bq{\begin{quote}}
\def\eq{\end{quote}}
\def\cbox#1#2 {\centerline{#1\fbox{#2}}}
\def\ul#1#2 {{#1\underline{#2}}}
\begin{document}
\title{\flushright{\small CERN-TH/99-337}\\
       \flushright{\small hep-ph/9911256}\\
Hard scattering in high-energy QCD$^\dagger$}

\author{Michelangelo L. Mangano$^\ddagger$}
%
\address{Theoretical Physics Division, CERN, 1211 Geneva 23,
  Switzerland\\
E-mail: {\tt michelangelo.mangano@cern.ch}}

\abstract{I review the recent results in the field of QCD at high
  energy presented to this Conference. In particular, I will
  concentrate on measurements of $\as$ from studies of event
  structures and jet rates, jet production in hadronic collisions, and
  heavy quark production.}

\maketitle
\fntext{\dag}{Rapporteur talk given at the 1999 EPS Conference,
  July 1999, Tampere, Finland}
\fntext{$\ddagger$}{This work was supported in part by
    the EU Fourth Framework Programme ``Training and Mobility of
    Researchers'', Network ``Quantum Chromodynamics and the Deep
    Structure of Elementary Particles'', contract FMRX--CT98--0194 (DG
    12 -- MIHT).} 

\section{INTRODUCTION}
There is no doubt nowadays that QCD {\em is} the theory of strong
interactions. Studies of QCD at high energy have left the phase of
{\em testing} QCD, and have stepped into the phase of accurate
measurements. A priori, the only fundamental quantity relevant to the
high-energy regime of QCD that needs to be measured is $\as$, as, in
principle, everything else can be calculated from first principles.
$\as$ is one of the fundamental ``constants'' of nature, and as such
there is no limit to the accuracy with which we would like to know its
value. The accurate knowledge of $\as$, for example, is necessary to
extract electroweak parameters from precision $e^+e^-$ data, as well
as to explore scenarios in which all forces of nature are unified at
some high-energy scale.  Pragmatically, one can therefore say that 
QCD studies at high energy mostly aim at assessing the accuracy of the
approximations used by theorists to evaluate QCD cross sections. This
accuracy forms the basis of our evaluation of the systematic errors
entering the extraction of $\as$ from the data. Confidence in our
practical understanding of QCD, on the other hand, has also an impact
on our ability to predict cross-sections for processes which produce
potential backgrounds to new physics, as well as to estimate the
potential of experiments at future accelerators. An accurate
understanding of high-energy QCD, furthermore, is necessary for an
accurate determination of several other important parameters of the
SM, such as the $W$ mass (whose measurement is limited theoretically
by the knowledge of structure functions in hadronic collisions, and of
parton recombination effects in $e^+e^-$), and the top mass (where
the properties of multi-gluon emission from initial and final states
limit the theoretical accuracy of determinations from hadronic
colliders). Upon closer scrutiny, one quickly realises that the ideal
world in which $\as$ is the only ingredient needed to perform precise
QCD predictions is far from reality. Since the observable final states
are made out of mesons and baryons, and not of quarks and gluons,
understanding of the interface between hard and soft QCD cannot be
escaped even when we consider high-energy processes. This is
particularly true when the initial states are given by hadrons, rather
than by lepton pairs. As a result, ancillary information (such as
parton densities and fragmentation functions) is needed for all
practical cases. Deviations from the ideal factorisation regime, where
all non-perturbative knowledge is dumped into phenomenological
quantities, become also relevant when dealing with observables which
have potentially large higher-twist (or power suppressed)
contributions. 

At the end of the day, one therefore concludes that future progress in
QCD phenomenology will still benefit enormously from all possible
sources of experimental information. It was then a great pleasure for
me to deal with the almost 100 papers submitted to the QCD session of
this Conference, number which testifies of the great interest which
the field still raises, and of the great progress made by the
experimental groups in developing new analysis tools and extracting
useful information from their rich data samples.  Given the large
number of papers submitted, it will be difficult however to review in detail
each contribution.  I will try to keep a balance between the in-depth
review of some critical issues, and the completeness of the report. As
a result, I hope to provide the reader with a sort of hitchhiker's
guide to the current open problems and to the available results,
hoping that this will stimulate some to tackle the most
interesting problems.

Since  most submissions reviewed here have not appeared
as yet as preprints, I have included references to the
number of the submitted papers in the form (X\_YYY), where X labels
the session and YYY the paper number. The write-ups of most of
these papers can be found from the Web pages of the experiments,
listed in refs.~\cite{ALEPH}-\cite{D0}.  To conclude this
Introduction, I wish to thank the QCD physics coordinators of all
experiments, for the prompt submission of draft papers, and for
replying to my enquiries. I also thank S. Catani for several
discussions related to the contents of this talk.

\section{EXPERIMENTAL INPUTS AND THEORY TOOLS}
The experimental papers submitted to the Conference cover all possible
aspects of high-energy QCD phenomenology. Issues related to the
structure of the proton and of the photon (parton densities,
diffractive phenomena, small-$x$ physics) are reviewed in the reports
by Marage and Barreiro. The set of inputs I will cover in this talk
is summarised here:
\begin{itemize}
  \item Properties of final states (shapes, multiplicities,
    fragmentation  functions,
    heavy-quark fractions). These have an important impact on perturbative QCD
    studies, as well as on the study of power corrections and
    non-perturbative physics, and on the extraction of $\as(M_Z)$
  \item Jet production in $p \bar p$, $e^+e^-$ and $ep$ collisions:
    these phenomena allow to explore the proton structure at the
    smallest possible distance scales, as well as to extract
    information on $\as(M_Z)$
  \item $W/Z$ production
  \item Heavy quark production: gluon splitting to $c\bar c$ and
    $b\bar b$, pair production in $\gamma\gamma$, $ep$ and
    $\gamma\gamma$ collisions.
\end{itemize}
Although several papers have been submitted, I will unfortunately be
forced to leave out for lack of time and space topics which are not
(or not yet) directly related to the regime of {\em hard} QCD:
detection and study of resonances inside jets, Bose-Einstein
correlations, Fermi-Dirac correlations, multi-particle production, etc.
These subjects were recently reviewed by B. Webber~\cite{Webber} in
his rapporteur presentation at the 1999 Lepton-Photon Symposium, and I
refer the readers to his forthcoming contribution to the Proceedings
for a good review of the topic.

The set of theoretical results presented at this conference represents
only a minor sample of the most recent developments. A common thread
among these contributions is the attempt to extend the range of
applicability of perturbation theory (PT), by either increasing the
order at which the perturbative calculations are performed, by
resumming classes of potentially large (logarithmic) contributions
appearing at all orders of PT, or by exploring the behaviour of PT in
the infrared domain in the attempt to provide a phenomenological
description of the hadronisation phase.

({\em i}) Techniques introduced recently~\cite{Draggiotis:1998gr}
for the evaluation of LO multi-particle amplitudes have
been reviewed here by Draggiotis~(1\_652).
({\em ii}) The achievement of NNLO accuracy for jet observables in hadronic
collisions represents today an open challenge for theorists. Several
milestones need to be met before NNLO cross-sections become available.
Among these milestones is the evaluation of the singular behaviour of
amplitudes near multi-collinear and multi-soft poles. Analytical
control over such behaviour allows to factorise the infrared and
collinear singularities which appear at higher orders of PT, and is an
important element in the construction of higher-order, parton-level
event generators. Important progress has occurred in the recent
past~\cite{Bern:1998sc}, as was discussed here by Uwer~(1\_129). 
({\em iii}) Harlander~(1\_444) presented the results of a recent
calculation~\cite{Chetyrkin:1999bx} of the
corrections of $\calO(\alpha_s^3,m_Q^4)$ to $R_{\rm had}$, a
quantity whose experimental measurement has reached high accuracy and
for which higher-order mass corrections are non-negligible.
({\em iv}) Progress in the resummation of NLL threshold corrections,
relevant for the production of heavy quarks, jets and prompt
$\gamma$'s at the edge of phase
space~\cite{Kidonakis:1997gm}-\cite{Catani:1999hs}, was
reviewed by Kidonakis~(1\_164)~\cite{Kidonakis:1999xm}. Some
applications in the case of top quark and prompt-$\gamma$ production
will be discussed in more detail later on.  ({\em v}) Sterman~(1\_712)
reviewed the status of the analytical understanding of power
corrections~\cite{Sterman:1999gz}, and ({\em vi}) explorations of the
PT-non-PT transition region in QCD were illustrated by Eden~(1\_206).

\section{SANITY CHECKS OF QCD}
A large number of results submitted to the Conference do not find an
obvious classification. To cope with this problem, I allowed myself to
introduce a new category, namely that of {\em sanity checks} of QCD.
I include in this category tests of QCD and QFT which do not have a
direct impact on specific measurements (or not yet!).  The lack of a
direct impact could be due to several reasons:
\begin{itemize}
  \item the measurements are not sufficiently accurate to improve our
  knowledge of some fundamental parameter (e.g. $m_b$)
  \item they are qualitative in nature (e.g. tests of colour coherence)
  \item they explore hard-wired fundamental features of QCD (e.g.  the
    flavour independence of $\as$, $N_c=3$, \dots), and any deviation
    from the expected result would not signal a problem with QCD
    itself, but most likely the failure of some approximation used in
    the theoretical calculations, or, ultimately, the presence of unexpected
    new physics.
\end{itemize}
By no means classification under ``sanity checks'' should be taken as
a negative judgement of the merits of a measurement.  The tests
discussed in the following testify in fact of the increased
sophistication of experimental techniques. Furthermore, in several
cases they set the groundwork for possible future applications, e.g.:
\begin{itemize}
  \item background removal in searches for New Physics
  \item use of $q$-jet versus $g$-jet discrimination 
  \item use of colour-coherence patterns to separate
  production of colour-singlet objects from multi-jet backgrounds
\end{itemize}

\subsection{Evolution with $\sqrt{S}$}
One of the most fundamental sanity checks of QCD is the
behaviour of observables w.r.t. changes in the hardness of the
process. The first preliminary measurements at the new energy frontier
of LEP ($\sqrt{S}=192-196$~GeV) have been submitted to this Conference
(Aleph: (1\_392); L3: (1\_232); OPAL: (1\_80)). There is no deviation from
the expected evolution of track multiplicities, momentum fractions,
and jet multiplicity rates.

\subsection{Quark-mass effects} 
Improved tagging techniques have led in the recent past to a
flourishing of QCD measurements based on the identification of heavy
quarks inside the jets. This allows to test characteristic properties
of Quantum Field Theory, such as the running of quark masses with
energy or the screening of collinear singularities induced by the
quark mass. In addition, as discussed later, the tagging of specific
flavours inside the jets allows for example to isolate samples of
high-purity gluon jets. 

I will start from measurements of the b-quark mass at $M_{Z^0}$. These
measurements use properties of $b$-tagged events such as the 3-jet
rate or moments of some shape variable (e.g. $B_{W,2}$, the second
moment of the wide-jet broadening). The $b$-quark mass $m_b$ appears
as a parameter in the theoretical evaluation of these
quantities~\cite{Bernreuther:1997jn}-\cite{Oleari:1997az}, and can be
fitted from the the data. It is usually more convenient to work
with a running $\MSB$ mass, in which case the value one expects to
extract from the fits to the data is the running $b$ mass,
evaluated at the scale $M_Z$. The results shown at this Conference are
collected here (values in GeV):
\\[0.2cm]
(1\_384) ALEPH:
  \ba m_b(M_Z) &=&
  \begin{array}{ll}
    3.04 \pm 0.92 & \mbox{3-jet fraction} \\
    3.78\pm 0.27  & B_{W_2} 
  \end{array}
\nn \\
m_b^{\MSB}(m_b) &=&
  \begin{array}{ll}
    4.16 \pm 1.10  & \mbox{3-jet fraction} \\
    5.04\pm 0.32  & B_{W_2}
  \end{array}
\nn   \ea
(1\_223) DELPHI~\cite{Abreu:1997ey}:
  \[ m_b(M_Z)=2.61\pm .18_{st}{+.45 \atop -.49}_{frag}
 \pm.04_{tag}\pm.07_{th}\] 
(1\_449) A. Brandenburg et al.~\cite{Brandenburg:1999nb}:
  \[ m_b(M_Z)=2.52\pm .27_{st}{+.33 \atop -.47}_{sys}
  {+.28 \atop -1.39}_{had}\pm .48_{th} \] 
These values are consistent with those obtained from the standard
determination of $m_b^{\MSB}(m_b)=4.20(8)$, obtained from the
application of NNLO PT and QCD sum rules to the $\Upsilon(1S)$
system~\cite{Melnikov:1998ug}. The systematic uncertainties are
however very large, and these measurements should therefore be taken
just as overall consistency checks. For a comparative study of the
measurements performed by Aleph and Delphi, see
ref.~\cite{Palla:1999st}.

The large value of the bottom quark mass is predicted by QCD to
affect also other observables, such as for example the multiplicity of
the final state.  Soft-gluon emission is suppressed within a cone (the
{\em dead} cone) of radius $2m_b/\sqrt{S}$ around the direction of the
$b$ quark, while it is identical to that originating from a light
quark outside the dead cone. The difference between the 
multiplicities of heavy- and light-quark final states is then
approximately proportional to the number of gluons emitted by the
light quark inside the dead cone area, corrected by the average
multiplicity produced during the heavy quark weak decay. When
$\sqrt{S}$ increases, the size of the dead cone diminishes, but the
density of emission from the light quark increases, and the two
effects compensate each other, leaving a constant multiplicity. As a
result, the difference of the average charged multiplicity of $b\bar
b$ and light-quark events as a function of $\sqrt{S}$ is a
constant~\cite{Khoze:1997dn}:
\[ \delta_{bl} \equiv \langle n\rangle_{bb} - \langle n\rangle_{ll} 
   \sim \mbox{const}\, .\]
This relation (valid asymptotically, and up to potentially large
corrections of $\calO(\sqrt{\as})$) is supported by the recent
measurements by DELPHI~(1\_220), which used the large
lever arm in energy available between LEP and LEP2. The
results~\cite{Muller:1999cy}  are
consistent with the theoretical expectations of a constant
$\delta_{bl}$, although the extracted value is slightly larger than
anticipated, indicating large sub-dominant contributions.
\[
\delta_{bl} = \begin{array}{ll}
  2.96\pm0.20 & \sqrt{S}=91.2~GeV \\
  5.07\pm1.28_{stat}\pm1.07_{syst} & \sqrt{S}=183~GeV \\
  3.97\pm0.83_{stat}\pm0.68_{syst} & \sqrt{S}=189~GeV \end{array}
\]

\subsection{Flavour independence of $\as$} 
These tests are performed by isolating samples of tagged heavy-flavour
events, and extracting the value of $\as$ from standard event-shape
and multi-jet analyses, using \oatwo\ QCD calculations including the
heavy-quark mass effects~\cite{Bernreuther:1997jn}-\cite{Oleari:1997az},
\cite{Ballestrero:1992ed}. The following recent results were submitted
to this Conference:
\\[0.2cm]
(1\_25) OPAL~\cite{Abbiendi:1999fs}: 
\[ \frac{\alpha_{\sss
    S}^{c}}{\alpha_{\sss S}^{f}} = 0.997\pm 0.050 \; , \;
\frac{\alpha_{\sss S}^{b}}{\alpha_{\sss S}^{f}} = 0.993\pm 0.015 
\]
 (1\_223) DELPHI:
\[ \frac{\alpha_{\sss S}^{b}}{\alpha_{\sss S}^{f}} =1.005\pm 0.012 \]
The flavour independence of $\as$ is therefore tested at the 5\% level
for charm, and at the \% level for bottom.

\subsection{$b$ couplings}
Final states where the $b\bar b$ pair is accompanied by
the emission of a hard gluon can be used to test the Lorentz structure
of the $b\bar b g$ coupling, through the study of the spectrum and angular
correlations of the 3 jets in the event. Limits on the anomalous
chromomagnetic coupling $\kappa$ defined by:
\[
\Delta{\cal L}=\frac{\kappa}{4m_b}g_s\,\bar
b\sigma_{\mu\nu}b\,G_{\mu\nu}
\]
have been obtained by SLD (1\_182): $-0.11 < \kappa <
0.08$~\cite{Abe:1999pz}.  In a further study SLD~(1\_183) analysed
correlations in the $b\bar b g$ decay of polarized ${Z^0}$. They
found no evidence for T-odd CP-even or T-odd CP-odd asymmetries at a
level of 5\%~\cite{Abe:1999qc}.

\subsection{Colour coherence}
Quantum coherence is a fundamental property of radiation from
multi-parton states. In the case of gluon emission, this is often
referred to as colour-coherence, and manifests itself with non-trivial
angular emission patterns for soft radiation~\cite{Dokshitzer:1991wu}.
Being a property of quantum theory, there is no doubt it should be
there! As a result, it could become a very interesting tool to learn
more about the underlying parton-level structure of a final state, for
example to separate quark and gluon jets, or to identify events where
jets come from colour-singlet sources (such as the decay of a neutral
object) from generic QCD events. The following studies were presented at
this Conference, providing further evidence that colour-coherence
effects survive the parton-to-hadron transition, and 
can be detected with appropriate selection criteria:
\\[0.2cm]
(1\_163e) D$\emptyset$: Evidence of color coherence in $W$+jets events in
$p\bar{p}$ at $\sqrt{s}$ = 1.8 TeV~\cite{Abbott:1999cu}.
\\{} 
(1\_145) DELPHI: A test of QCD coherence and LPHD using 
symmetric 3-jet events. \\
{}(1\_510) DELPHI: Testing of the New Parton Final State
Reconstruction Method Using $Z^0\to b\bar{b}g$ Mercedes Events.

\subsection{$C_A/C_F$ from $N_{ch}$ and $F(z)$ in $q/g$ jets}
\begin{figure}
\centerline{
 \includegraphics[width=0.45\textwidth,clip]{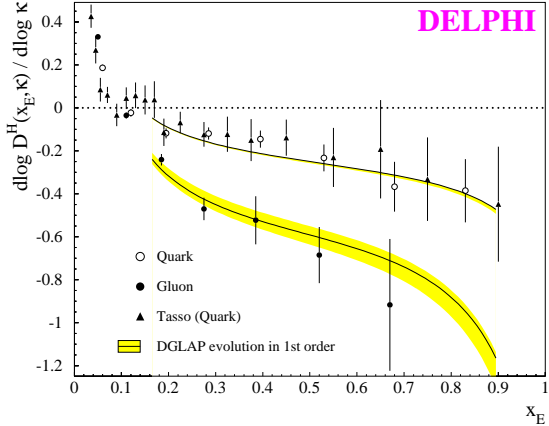}}
\vspace*{-0.5cm}
\caption{Scaling violations in the fragmentation functions $D(z)$ for quark
 and gluon jets, from DELPHI.}
\label{fig:delphi_scavio}
\end{figure}

Heavy-flavour tagging techniques have also found a useful application
in the study of fragmentation properties of light partons. The tagging
of b quarks in a 3-jet event allows in fact to single out the gluon
jet, and anti-tagged events provide samples of light-quarks. Using
events with different 3-jet kinematics, it is possible to probe the
jet properties at different scales, and test the scale evolution of
observables such as the jet fragmentation function and the jet
charged multiplicity $N_{ch}$. 
QCD provides detailed predictions for the evolution of
these quantities and predicts clear differences between quark and
gluon jets. These differences have been now measured with great
accuracy, and have been turned into a measurement of the relative
colour charge of gluons and quarks ($C_A/C_F$, equal to 9/4 in QCD).
The following is a summary of these most recent results. A more
complete account can be found in the contribution to these
Proceedings by  B.~Gary~\cite{Gary:1999bp}.
\\[0.2cm]
(1\_571) DELPHI: Gluon fragmentation function and scaling
violations in q/g jets (see fig.~\ref{fig:delphi_scavio}):
\[ 
\frac{C_A}{C_F} = 2.23(9)(6)
\]
\\
(1\_383) DELPHI:
 Scale Dependence of $N_{ch}$ in $q$ and $g$ jets~\cite{Abreu:1999rs}:
\[
\frac{C_A}{C_F} = 2.246 \pm 0.062_{stat} \pm 0.080_{sys} \pm 0.095_{th}
\]
(1\_24) OPAL:  Experimental properties of gluon and quark 
      jets from a  point source~\cite{Abbiendi:1999pi}:
\[
\frac{C_A}{C_F} = 2.29\pm0.09_{\mathrm{stat}}\pm0.15_{\mathrm{syst}}
\]
(See also OPAL~(1\_6), ``A simultaneous measurement of $\as$ and QCD
colour factors'').  Interesting comparisons of the multiplicity of
individual hadrons in quarks and gluon jets were presented by
OPAL~(1\_4) and DELPHI~(3\_146). OPAL found that 
the observed ratio of $\eta$ mesons
in quark and in gluon jets agrees with what predicted by QCD Monte
Carlo calculations, while the absolute number of $\eta$'s is larger in
the data than in the MC. Similar small discrepancies in the absolute
prediction for kaons and protons in gluon jets have been observed by DELPHI.
More details can be found in ref.~\cite{Webber}.

\section{$\as(M_Z)$ MEASUREMENTS}

\subsection{Event shapes in $e^+e^-$ and $ep$}
Event shapes are defined as infrared and collinear safe observables,
characterising the structure of the final state. By construction, they
should be calculable in PT, up to finite non-PT corrections suppressed
by powers of the typical hard scale of the process. They are sensitive
to properties of QCD radiation, and therefore allow the measurement of
$\as$.  QCD  predictions are available at
$\calO(\astwo)$~\cite{Ellis:1981wv,Kunszt:1989km,Catani:1997vz}, which in the
case of $e^+e^-$ and of DIS corresponds to the NLO in PT. In most
cases, these NLO calculations are improved by the resummation of
next-to-leading logarithms (NLL) appearing at all orders of PT near
the edge of phase-space~\cite{Catani:1993ua}.  Since shape variables are sensitive to
non-perturbative power-suppressed (i.e. $ \propto 1/Q $) effects,
their study allows the analysis of the hadronisation phase.

Extractions of $\as$ in the past have used a description of non-PT
corrections to event-shapes based on the hadronisation models of
shower MC's (Herwig, Jetset). With the recent progress in the
theoretical understanding of the structure of power corrections
\cite{Manohar:1995kq} the
effect of hadronisation corrections can be described
with simple analytical expressions, dependent on a small number of
parameters.
These parameters, in turn, can be interpreted~\cite{Dokshitzer:1996qm}
as moments of the strong coupling constant, averaged over the small
$Q^2$ region. For most cases, what is required is the first moment of 
$ \as(Q)$:
\be
 \alpha_0(\mu_0) = \frac{1}{\mu_0}\; \int_0^{\mu_0} \, {\rm d}q \;
 \as(q)
\ee
(in practical cases, one usually takes $\mu_0=2$~GeV).
Early experimental tests of these ideas were presented in
refs.~\cite{Abreu:1996mk}-\cite{Biebel:1999zt}, where evidence was
found for values of $\alpha_0(2\mbox{GeV})\sim 0.5$.

It is common to use either {\em moments} of shape variables, or {\em
 distributions}. 
In the case of the first moments:
\be \langle \calF \rangle = \langle \calF_{PT} \rangle +\langle \calF_{non-PT}
\rangle 
\ee
where $\langle \calF_{PT} \rangle$ is obtained from the PT
 calculation, while the non-PT effects are shown to factorise into the
 contribution $\langle \calF_{non-PT} \rangle = c_F \; \calP$. Here
\be \calP =
\frac{4C_F}{\pi^2}\,\calM\,\frac{\mu_0}{\sqrt{S}}\,\left[\alpha_0(\mu_0)
- \as(\sqrt{s})+\calO(\astwo)\right] \, .
\ee
is a universal factor ($\calM=1.795$ is the so-called ``Milan'' 
factor~\cite{Dokshitzer:1997iz}, incorporating effects of 2-loop
 corrections), and $c_F$ is an observable-dependent coefficient:
\be
 { \begin{array}{clllll}
 \calF= & 1-T & M^2_H & B_T & B_W & C \\
  c_F =& 2 & 2 & 1 & 1/2 & 3\pi
  \end{array}}
\ee

In the case of  distributions, the effect of power corrections is to
shift the value of the observable in the PT QCD
prediction~\cite{Korchemsky:1995is} (so long as the distance of the
observable from its kinematic threshold value is larger than $\calO(1/Q)$):
\be \frac{d\sigma}{d\calF}(\calF) = 
    \frac{d\sigma^{PT}}{d\calF}(\calF-\calP \, D_{\calF}) 
\ee
$\calP$ is the same as for the 1st moment, and $D_{\calF}$ can
be calculated for each observable. It can be a constant:
\be
{D_{\calF}=\left\{ \begin{array}{ll}
  2 & \calF=1-T \\
  3\pi & \calF=C \end{array} \right. } 
\ee
or a function, as recently shown in the case of jet broadenings in
ref.~\cite{Dokshitzer:1998qp} ($\calF=B_T, \, B_W $):
\be \label{eq:bdepshift}
D_\calF =\frac{1}{2}\log\frac{1}{\calF}\,+\,
B_{\calF}(\calF,\as(\calF\sqrt{s})) 
\ee

\subsection{Shape variables: results from QCD+MC fits}
Let us begin with $\as(M_Z)$ determinations obtained from NLO+NLL QCD fits
with non-PT effects described via MC programs. The following new
measurements have been reported at this Conference:
\\[0.2cm]
(1\_410) ALEPH (the third value, labelled by a (*), refers to the
  combination of LEP1+LEP2 results):
\[ \begin{array}{rcl}
 \as(189) &= &0.1119(15)_{stat}(11)_{sys}(30)_{th} \\
 \as(M_Z) &= &0.1249(44)\\
  \alpha_s^{*}(M_Z) &= & 0.1216(39) \end{array}\]
(1\_144) DELPHI (statistical errors only):
\[ \begin{array}{rcl}
 \as(189) &= &0.1116(24)_{stat}\\
 \as(M_Z) &= & 0.1246(30)_{stat}
   \end{array}\]
(1\_279) L3 (the value labelled by a (*) refers to the
  combination of results in the range 30--189~GeV):
\[ \begin{array}{rcl}
 \as(189) &= &0.1101(18)_{exp}(56)_{th} \\
 \as(M_Z) &= &0.1227(22)_{exp}(69)_{th}\\
    \alpha_s^{*}(M_Z) &= &  0.1220(62) \end{array} \]
(1\_5) OPAL (the value labelled by a (*)
uses data with $\sqrt{S}$ in the range 35--189~GeV, from JADE 
in addition to OPAL LEP1 and LEP2):
\[ \begin{array}{rcl}
 \as(189) &=
  &0.1085(15)_{stat}(27)_{sys}(20)_{had}\\ && \left({\sst+22\atop\sst-3}
  \right)_{scale}  \\
 \as(M_Z) &= &0.1206\left({\sst+54\atop\sst-46}\right)\\
  \alpha_s^{*}(M_Z) &= & 
  0.1199\left({\sst+38\atop\sst-25}\right)
\end{array}\]
An overall average of the above 189~GeV results, gives
\[\as(189)=0.1105(4), \quad \as(M_Z) =0.1232(5) \]
with the theoretical and systematic errors averaged among the various
experiments.

\subsection{Shape variables: results from QCD+$\calO(1/Q)$ fits}
\begin{figure}
\centerline{\includegraphics[bb=195 162 403
  412,width=0.5\textwidth,clip]{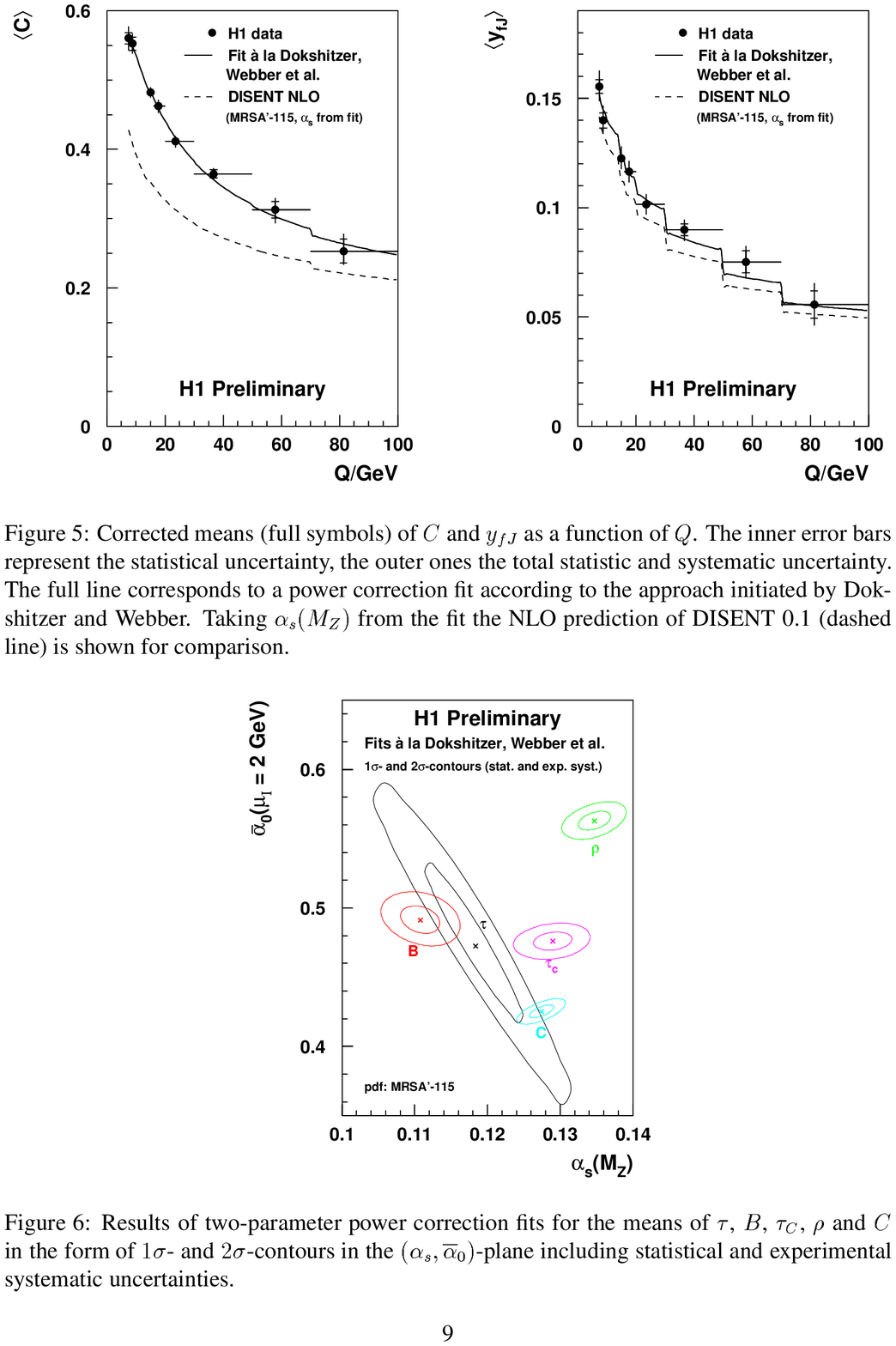}}
\vspace*{-0.5cm}
\caption{H1 fits for $\as(M_Z)$ and $\alpha_0(2\mbox{GeV})$, from
  different shape variables.}
\label{fig:h1a0fit}
\end{figure}
\begin{figure}
\centerline{\includegraphics[width=0.5\textwidth,clip]{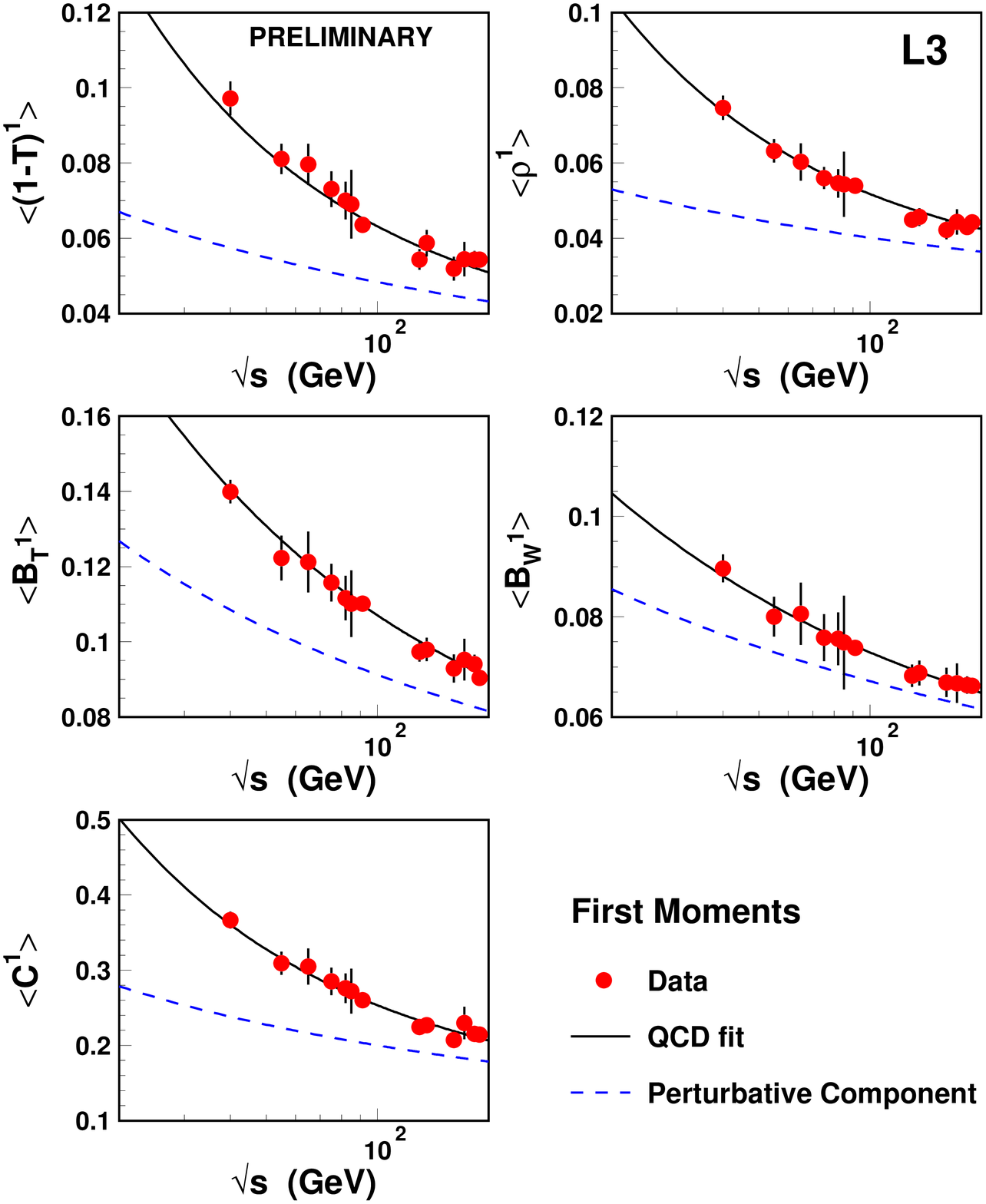}}
\vspace*{-0.5cm}
\caption{Evolution with $\sqrt{S}$ of the 1st moments of various
  shape variables, from L3. The dashed lines include only the
  PT contribution, the solid line is a fit including $\calO(1/Q)$
  effects. }
\label{fig:l3_shapes}
\end{figure}
\begin{figure}
\centerline{\includegraphics[width=0.5\textwidth,clip]{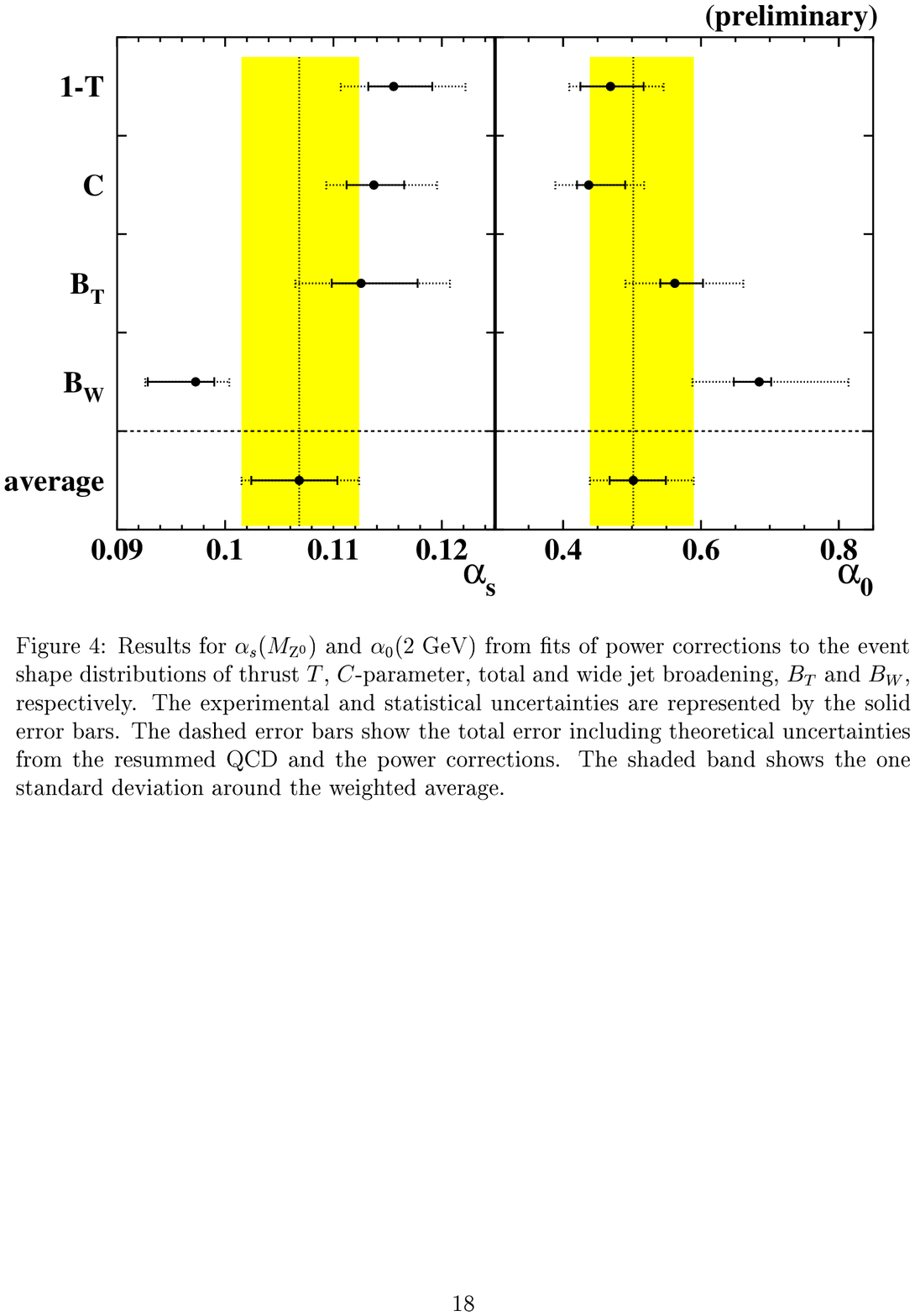}}
\vspace*{-0.5cm}
\caption{Fits for $\as(M_Z)$ and $\alpha_0(2\mbox{GeV})$, from
  different shape variables in $e^+e^-$ collisions 
  (Movilla~Fern\'andez et al.~\cite{MovillaFernandez:1999yn}).}
\label{fig:bethke_a0fit}
\end{figure}

We review here the recent $\as(M_Z)$ determinations obtained from QCD fits,
with non-PT effects described by analytic power corrections. 
\\
{\bf From 1$^{st}$ moments}: \\[0.2cm]
(1\_144) DELPHI: 
\[ \begin{array}{rcl}
 \alpha_0(2\,{\rm GeV}) &=& 0.5(1) \\
 \as(189) &=& 0.1102(23)_{stat}(18)_{sys}(24)_{th} \\
  \as(M_Z) &=&  0.1229(28)_{stat}(22)_{sys}(31)_{th}
 \end{array} \]
(1\_157k) H1 (Fig.~\ref{fig:h1a0fit}):
\[ \begin{array}{rcl}
 \alpha_0(2\,{\rm GeV}) &=& 0.50(5) \\
  \as(M_Z) &=&  0.12(1)\end{array} \]
{\bf From shape distributions:} \\[0.2cm]
(1\_279) L3 (Fig.~\ref{fig:l3_shapes}):
\[ \begin{array}{rcl}
 \alpha_0(2\,{\rm GeV}) &=& 0.490(46)\\
  \as(M_Z) &=&  0.1106(36)_{exp}(40)_{th}\end{array}  \]
(1\_113) Movilla~Fern\'andez et al, 
  ref.~\cite{MovillaFernandez:1999yn}, using data from 
 JADE and LEP (the value labeled by a (*) excludes the $B_W$
  distribution from the fit, see Fig.~\ref{fig:bethke_a0fit}):
\[ \begin{array}{rcl}
 \alpha_0(2\,{\rm GeV}) &=& 0.50{\sst+.9\atop\sst-.6} 
\\
  \as(M_Z) &=&  0.1068\pm.0011_{stat}{\sst+.0033\atop\sst-0.043}_{sys}
 \\ && {\sst+.0043\atop\sst-0.029}_{th}\\
  \alpha_s^{*}(M_Z) &=&  0.1141\pm.0012_{stat}{\sst+.0034\atop\sst-0.024}_{sys}
 \\ && {\sst+.0055\atop\sst-0.041}_{th}\end{array} \]

\begin{figure}
\centerline{ \includegraphics[width=0.5\textwidth,clip]{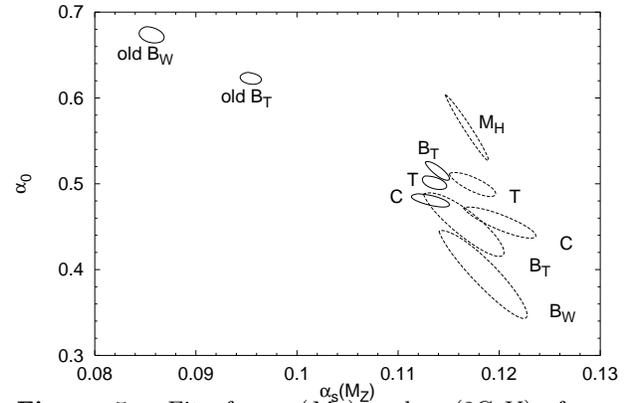}}
\vspace*{-0.5cm}
\caption{Fits for $\as(M_Z)$ and $\alpha_0(2\mbox{GeV})$, from
  different shape variables in $e^+e^-$ collisions. The dashed lines
  correspond to fit 1st moments, the solid lines to fits to
  distributions. Curves labeled by ``old'' refer to constant shifts in
  the broadening variable, while the others are obtained with the
  $B$-dependent shifts as in eq.(~\ref{eq:bdepshift}).
  (Dokshitzer et al.~\cite{Dokshitzer:1998qp}).}
\label{fig:salam_a0fit}
\end{figure}

All of the above measurements show a good consistency in the
extraction of $\as(M_Z)$ from QCD+hadronisation corrections, and from
1$^{st}$ moments using analytic power corrections. Significant
differences in $\as(M_Z)$ are vice-versa present when fitting shape
distributions with power corrections. This is particularly true for
the broadenings, for which the shift in the variable is given by the
function in eq.(~\ref{eq:bdepshift}). The inclusion of the variable
shift, due to a subtle interplay between perturbative and non-PT
effects, gives rise by itself to a large correction relative to the
naive prescription of a constant shift, as shown in
ref.~\cite{Dokshitzer:1998qp} and as displayed in
fig.~\ref{fig:salam_a0fit}. There are indications however that this
correction is not sufficient, and that a bigger {\em squeezing} in the
theoretical predictions is required. This is supported by the
comparison with hadronisation corrections predicted by
MC's~\cite{Biebel:1999zt}. I would conclude that we are moving in the
right direction for a phenomenological understanding of power
corrections, but more work is necessary before extractions of $\as$
can be improved further. It also remains to be evaluated whether the
breakdown of factorisation for power
corrections, shown in ref.~\cite{Nason:1995hd} to occur at higher orders, will
ultimately lead to an intrinsic limit in the theoretical accuracy
attainable with this approach.

\subsection{Other $\as(M_Z)$ measurements}
\begin{figure}
\centerline{
\includegraphics[width=0.4\textwidth,clip]{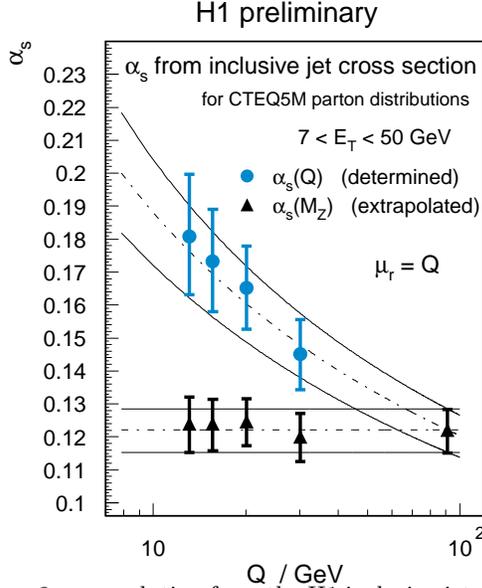}}
\vspace*{-0.5cm}
\caption{$\as$ evolution from the H1 inclusive jet rates.}
\label{fig:h1_asrun}
\end{figure}

\begin{figure}
\centerline{
\includegraphics[width=0.5\textwidth,clip]{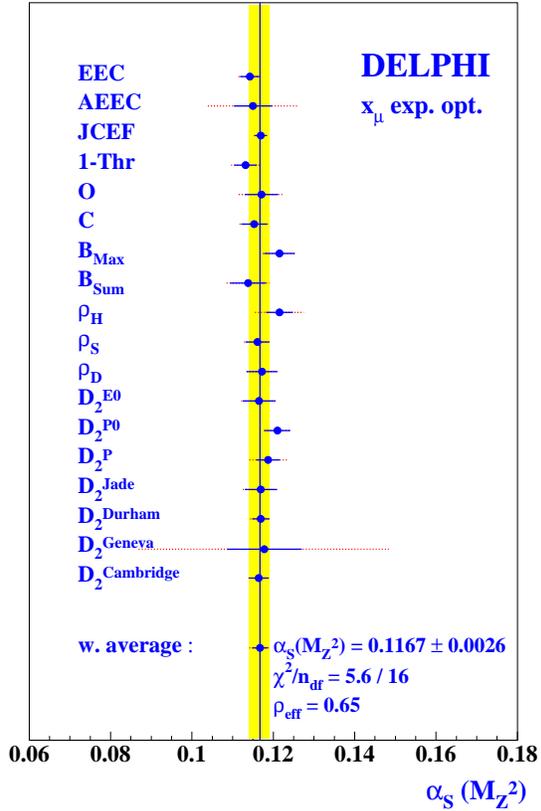}
}
\vspace*{-0.5cm}
\caption{Values for $\as(M_Z)$ extracted by DELPHI by optimising the
  scale choice on a variable-by-variable basis.}
\label{fig:delphi_asopt}
\end{figure}

In addition to the above results on $\as$ obtained from precise
shape-variable fits in $e^+e^-$ collisions, other measurements of
$\as$ have been reported at this Conference. The HERA measurements
have been reviewed by T.~Carli in his
presentation~\cite{Carli:1999eg}. I summarise here the main results:
 \\[0.2cm]
(1\_157) H1: Fit of the
Inclusive Jet Rate ${\rm d}^2\sigma/{\rm d}E_T{\rm d}Q^2$. The rates
are compared to NLO QCD
calculations~\cite{Catani:1997vz,Mirkes:1996ks}, 
and fit to the value of $\as$.
A recent comparison of various NLO
calculations for jet production in DIS can be found in
ref.~\cite{Duprel:1999wz}.
(see
fig.~\ref{fig:h1_asrun}; the first value below corresponds to
$\mu_R=E_T$, the second to $\mu_R=Q$):
\[
\as(M_Z)= \begin{array}{l}
 0.1181(30)_{exp}\left({\sst +39\atop\sst-46}\right)_{th}
                 \left({\sst +36\atop\sst-15}\right)_{PDF} \\
 0.1221(34)_{exp}\left({\sst +54\atop\sst-59}\right)_{th}
                 \left({\sst +33\atop\sst-16}\right)_{PDF} 
 \end{array}
\]
(1\_157y) H1: Fit of the Dijet Rate ${\rm d}n_{\rm dijet}/{\rm
  d}y_2$ (the first value corresponds to use of the Durham algorithm, in
  the lab frame, the second to the $k^{DIS}_{\perp}$
  algorithm in the Breit frame):
\[
\as(M_Z)= \begin{array}{l}
 0.1189\left({\sst +64\atop\sst-81}\right)_{exp}
       \left({\sst 59\atop\sst46}\right)_{th}
       \left({\sst 13\atop\sst55}\right)_{PDF}
        \\
\\
 0.1143\left({\sst +75\atop\sst-89}\right)_{exp}
       \left({\sst 74\atop\sst64}\right)_{th}
       \left({\sst 8 \atop\sst54}\right)_{PDF} 
\\
 \end{array}
\]
(1\_543) ZEUS: Fit of the Dijet fraction vs $Q^2$:
\[
\as(M_Z)= 0.120(3)_{stat}\left({\sst +5\atop\sst-6}\right)_{exp}
\left({\sst +3\atop\sst-2}\right)_{th}
\]
(1\_232) L3: QCD studies at 192 and 196~GeV:
\[ \begin{array}{rcl}
 \as(192) &=
  &0.1108(35)_{exp}(56)_{theory}  \\
 \as(M_Z)&=
  &0.1220(15)_{exp}(60)_{theory}
\end{array}\]
(1\_80) OPAL: QCD studies at 192 and 196~GeV (the value at $M_Z$
includes results in the range 30--192~GeV):
\[ \begin{array}{rcl}
 \as(193) &=
  &0.1025(38)_{stat}(54)_{syst}  \\
 \as(M_Z) &=
  &0.1135(47)_{stat}(67)_{syst}
\end{array}\]
(1\_2) OPAL: QCD studies at 172-189~GeV (the value at $M_Z$
includes results in the range 30--189~GeV):
\[ \begin{array}{rcl}
 \as(187) &=
  &0.106(1)_{stat}(4)_{syst}  \\
 \as(M_Z) &=
  &0.117(5)
\end{array}\]
(1\_157y) DELPHI: $\calO(\astwo)$ fits to oriented shape variables at
91.2~GeV: 
\[
\as(M_Z)= \begin{array}{ll}
 0.1173(23)&
       \mbox{18 shape variables}
\\
 0.1180(18) & 
              \mbox{Jet cone E fraction\cite{Ohnishi:1994vp}}
 \end{array}
\]
In this analysis~\cite{DelphiOSV}, the renormalisation scale is
``experimentally optimised'', i.e. it is chosen for each variable by
searching for the best fit to the relative data distribution. It is
extremely intriguing that the values of $\as$ extracted from each one of 18
variables are in extremely good agreement with one another (see
fig.~\ref{fig:delphi_asopt}), leading to the very small errors quoted
above. While this convergence is  tantalising, it must be
said that such an optimisation procedure has no theoretical basis, and
I personally do not consider therefore the  error estimate
theoretically solid.  Notice that there is a very large spread in the
resulting values of optimised scales~\cite{DelphiOSV} which go from
$x_{\mu}\equiv\mu^2/S=3\times 10^{-3}$ to $x_{\mu}=7$.  Altogether, I
have no precise understanding of which bias could cause such an
amazing convergence in the values of $\as$. The result is extremely
interesting, and certainly deserves theoretical attention, to try
understand whether such a bias exists, or whether the procedure itself
can be eventually justified. Until this understanding is achieved, I
would not advocate taking the quoted error literally when
 this measurement is included in global averages.

\subsection{$\as(M_Z)$ global averages}
The latest pre-1999 compilation and average of results on $\as$ was
given by S.  Bethke~\cite{Bethke:1998ja}.  Using measurements with
$\Delta\as<0.008$ only, he obtained: $ \alpha_S(M_Z) = 0.119\pm0.004
$.  If the values from Lattice QCD, which tend to have some of the
smallest absolute errors, are left out, then the average becomes:
$\alpha_S(M_Z) = 0.120\pm0.005$.

I don't dare stealing from S.Bethke the pleasure to produce the new
World Average for $\as(M_Z)$! Proper averaging of the new LEP results
will require detailed knowledge of the correlation matrices for the
various experiments, and will be done soon, I expect, by the QCD LEP
Working Group.  However, from the results submitted to this Conference
I don't see indications that the most recent updates on the value of
$\as$  will change significantly the
central value and the determination of the error on $\as(M_Z)$.  The
most recent extractions of $\as$ from the fits to jet shapes at
189~GeV support a slightly larger value of $\as(M_Z)$ relative to the
pre-EPS World average of $0.120\pm0.005$.  So my best bet for the next
world average is: $\as(M_Z)=0.121\pm 0.004$.

Progress in the analytic, phenomenological understanding of power
corrections, needed to extract $\as$ from jet shapes, is remarkable.
However, the current results should be taken in my view more as an
indication that the direction of these theoretical developments is
correct, rather than as a strong input for a reduction of the
theoretical error on $\as$.

Even more interestingly, they set the stage for future progress in the
area of jet physics in hadronic collisions, where large statistics and
huge lever arms in energy will lead to minuscule statistical
uncertainties on $\as$ in the future years. 

\section{ JET STRUCTURES IN $p\bar p$ and $ep$}
\begin{figure}
\centerline{
  \includegraphics[bb=77 319 357 672,
         width=0.35\textwidth,clip]{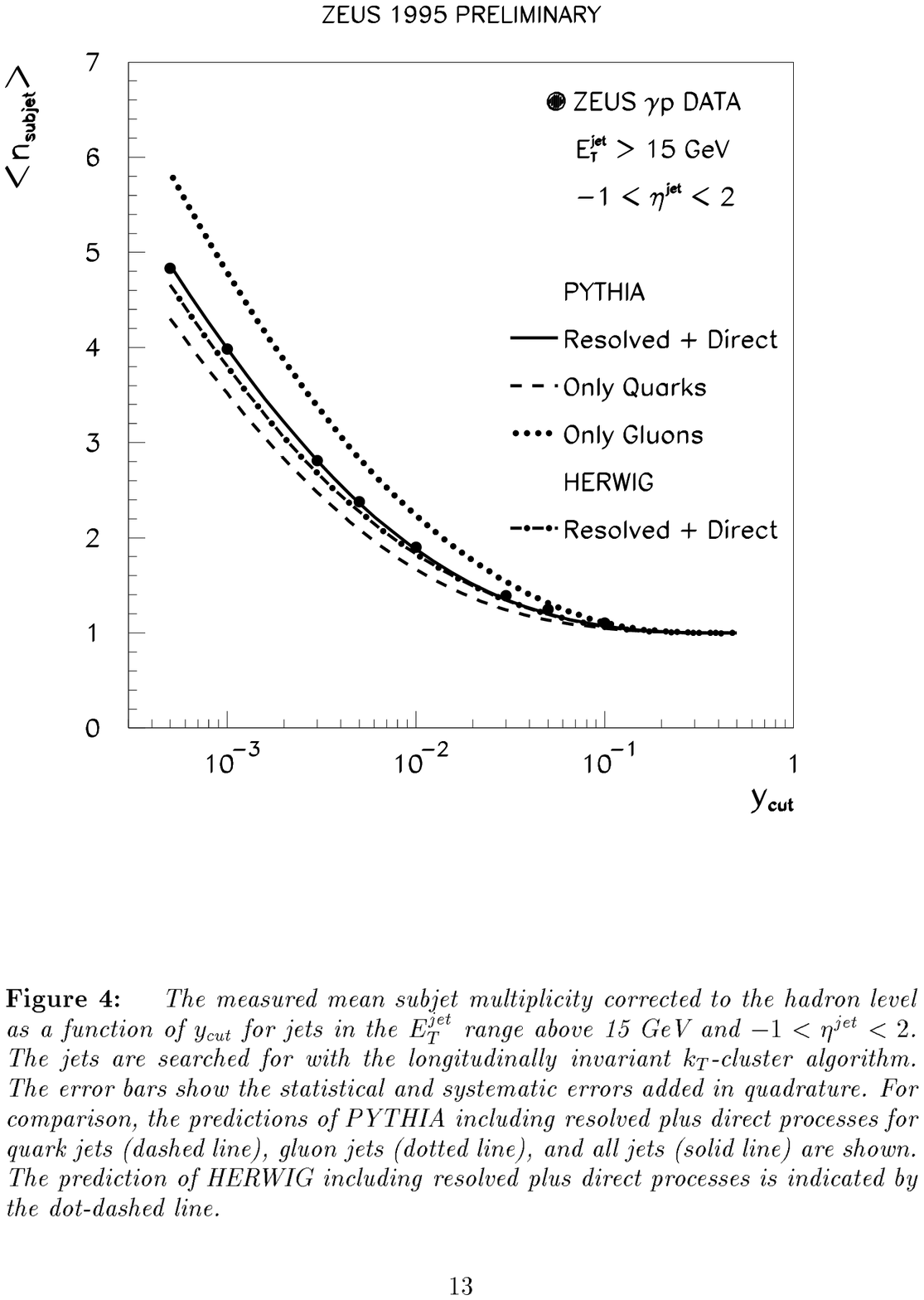}\hfill}
\vspace*{-0.5cm}
\caption{ZEUS results for the sub-jet multiplicity.}
\label{fig:zeus_subjet}
\end{figure}

\begin{figure}
\centerline{
  \includegraphics[width=0.45\textwidth,clip]{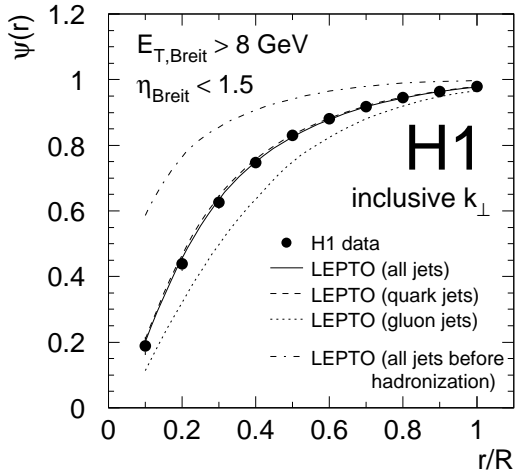}}
\vspace*{-0.5cm}
\caption{H1 distribution for the energy flow inside a jet.}
\label{fig:h1_subjet}
\end{figure}

Interesting new studies of the structure of jets in collisions
involving hadrons in the initial state have been presented. All
results confirm a good level of understanding of the theory of jet
substructure, although it is still premature to employ these studies
for accurate measurements of $\as$. \\[0.2cm]
(1\_163d) D$\emptyset$: Sub-jet multiplicity at $\sqrt{S}=630$ and
1800~GeV~\cite{Abbott:1999gj}. 
Data at the different energies, where the relative q/g
composition of jets are different for a fixed 
$E_T$, can be used to extract the average
number of sub-jets contained in a quark and in a gluon jet. The result
of the analysis gives:
\[ \frac{\langle n_j -1 \rangle _g}{\langle n_j -1 \rangle _q}=1.9\pm
0.2 \]
which is consistent with the LO QCD prediction of 9/4.
A theoretical analyses of sub-jet multiplicities in hadronic collisions
can be found in ref.~\cite{Forshaw:1999iv}.
\\{}
(1\_600) CDF: Jet fragmentation studies at the Tevatron~\cite{Safonov:1999dr}:
the distribution of $\xi=log(1/x)$ is studied for jets over a wide
range of energies, showing excellent agreement with MDLA both in terms
of shapes, and in the evolution of the peak position with energy.\\{}
(1\_530) ZEUS: Jet substructure in $\gamma$p. This is a study of the
sub-jet multiplicity as a function of the resolution parameter
$y_{cut}$. The distribution of the average $\langle n_j \rangle$ is
not consistent with the presence of a single type of partons
(either quarks or gluons), but is well fitted by the appropriate
composition of final state partons predicted by QCD, as shown in
fig.~\ref{fig:zeus_subjet}.
\\[0.2cm]
(1\_157x) H1: Jet substructure in  DIS
dijets. Figure~\ref{fig:h1_subjet} shows the fraction $\psi(r)$ of jet energy
contained in a sub-cone of radius $r$ inside the jet. As for the ZEUS
measurement presented above, the data are consistent with the right
composition of quark and gluon jets expected from QCD. 

\section{JETS AT THE TEVATRON}
At the Tevatron, jets up to 450 GeV transverse momentum have been
observed~\cite{cdfjets,Abbott:1998ya}. These data can be used for many
interesting purposes:
\begin{itemize}
\item Tests of QCD: calculations are available up to  NLO~\cite{jetsnlo}.
\item Extract information on the partonic densities, 
  $f_{q,g}(x,Q^2)$ at large $Q^2$.
\item Look for deviations from QCD (e.g. resonances in the dijet mass
  spectrum),  explore quark structure at small distances.
\end{itemize}
The results of the most recent analyses of the data from the full
Run~I of the Tevatron have been submitted to this
Conference~\cite{Gallas:1999yp} 
(D$\emptyset$: (1\_163c); CDF: (1\_593) and (1\_594)).

\begin{figure}
\begin{center}
\includegraphics[width=0.5\textwidth,clip]{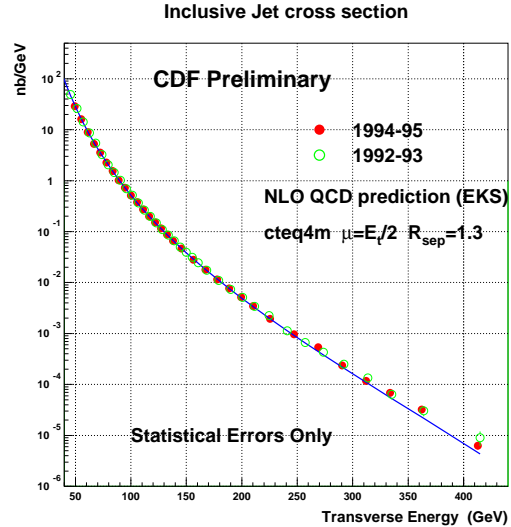}
\vspace*{-0.5cm}
\caption{Inclusive jet transverse energy ($\et$)
distribution as measured by CDF, compared to the absolute NLO QCD
calculation.}
 \label{fig:cdfjet} 
\end{center}
\end{figure}

\begin{figure}
\begin{center}
\includegraphics[width=0.5\textwidth,clip]{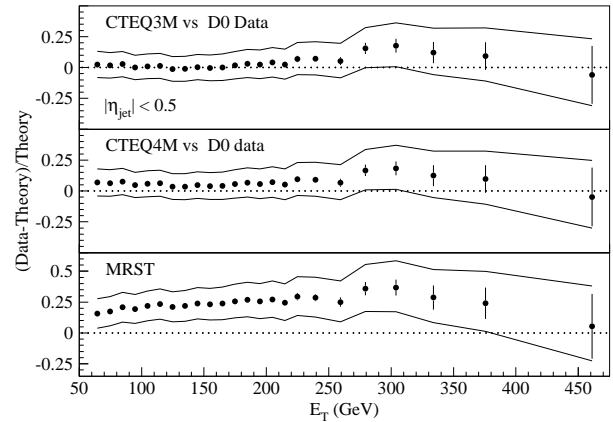}
\vspace*{-0.5cm}
\caption{Deviations of QCD predictions from  D$\emptyset$ jet
data for various sets of PDFs.}
\label{fig:d0jet_pdf}
\end{center}
\end{figure}

The studied range of transverse energies corresponds to values of
$x \gappeq 0.5$, at $Q^2 \simeq 160,000$ GeV$^2$.  This is a domain
not accessible to DIS experiments.  The
current agreement between theory and data is at the level of 30 \%
over 8 orders of magnitude of cross-section, from $E_T\sim$ 20 to
$E_T\sim$ 450 GeV (see fig.~\ref{fig:cdfjet})
In spite of the general good agreement, a large dependence on the
chosen set of parton densities~\cite{Martin:1998sq,Lai:1999wy}
 is present, as shown in
fig.~\ref{fig:d0jet_pdf}. The presence of this uncertainty limits the
use of high-$\et$ jet data to set constraints on possible new physics.

An important question is therefore the following: to which extent do
independent measurements of parton densities constrain the knowledge
of PDFs at large $x$, and what is the residual uncertainty on the jet
$\et$ distributions?

\begin{figure}
\begin{center}
\includegraphics[width=0.5\textwidth,clip]{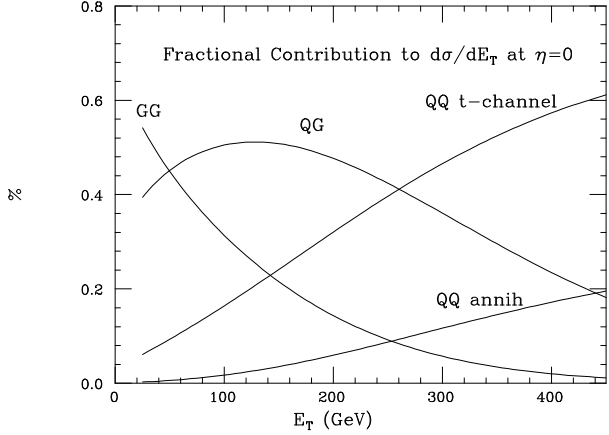}
\vspace*{-0.5cm}
\caption{Contributions from different initial states to the jet cross
  section at $\sqrt{s}=1.8$~TeV}
\label{fig:isfrac}
\end{center}
\end{figure}
To address this issue, let us first show what is the relative
contribution of different initial state partons to the jet cross
section. This is plotted in fig.~\ref{fig:isfrac}, where some
standard PDF set (CTEQ4M~\cite{Lai:1999wy} in this case) was chosen.
At the largest energies accessible to today's Tevatron data, 80\% of
the jets are produced by collisions involving only initial state
quarks. The remaining 20\% comes from processes where at least one
gluon was present in the initial state.

Quark densities at large $x$ are constrained by DIS data to 
within few percent, leading to an overall uncertainty on the
high-$\et$ jet rate of at most 5\%. 
What is the uncertainty on the remaining 20\% coming from gluon-induced
processes?
How are we guaranteed that the gluons are known to better than a
factor of 2, limiting the overall uncertainty to 20-30\%?

\begin{figure}
\begin{center}
\includegraphics[width=0.5\textwidth,clip]{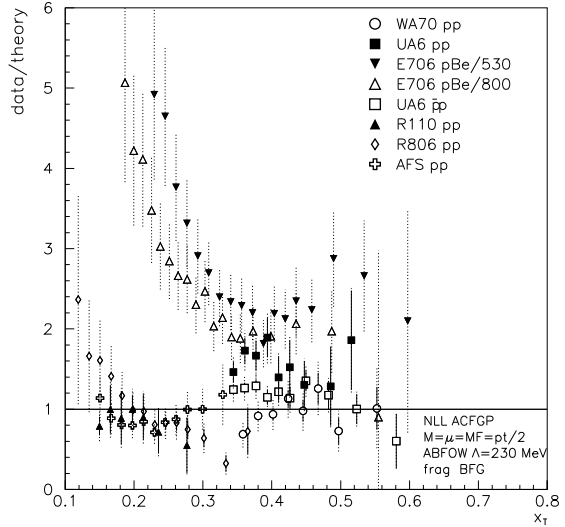}
\vspace{-.7cm}
\caption{ Relative deviations between NLO
QCD and prompt photon data, as a function of $x_T=2\pt/\sqrt{S}$, for
various fixed target experiments.}
\label{fig:aurenche_allxs}
\end{center}
\end{figure}

\begin{figure}
\begin{center}
\includegraphics[width=0.5\textwidth,clip]{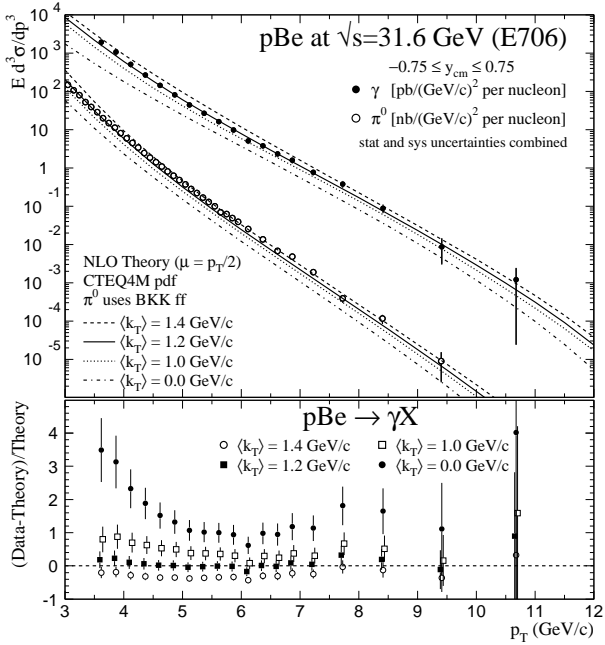}
\vspace{-0.5cm}
\caption{ Comparison of E706 data~\cite{Apanasevich:1998hm} with NLO QCD,
before and after inclusion of an intrinsic-$k_T$.}
\label{fig:tung3}
\end{center}
\end{figure}

The only independent constraint on $ f_g(x,Q^2)$ comes from
fixed-target production of prompt photons. This process is induced at
LO by two mechanisms, $q\bar q \to g \gamma$ and $qg \to q \gamma$.
In $pN$ collisions $g(x)\gg \bar{q}(x)$, and therefore
${d\sigma}/{dE_T}(qg\to q\gamma) \gg    {d\sigma}/{dE_T}(q\bar q\to g\gamma) $
Data from FNAL and CERN fixed target experiments can therefore be used to
extract $ f_g(x,Q^2)$ at large $x$. 
Unfortunately, a comparison~\cite{Aurenche:1998gv} of data and NLO theory shows
discrepancies at small
$E_T$, as well as inconsistencies between the various experiments, as shown in
fig.~\ref{fig:aurenche_allxs} (1\_635).
The authors of ref.~\cite{Aurenche:1998gv} suggest that the large scale
uncertainty in the theoretical calculations can be sufficient to
explain away the differences observed at small $E_T$ between data and
theory, in particular in view of the apparent inconsistency between
some of the experimental results.
As a possible additional explanation for these discrepancies, the presence of a
large non-perturbative contribution from the intrinsic $k_T$ of
partons inside the nucleon has been
suggested~\cite{Apanasevich:1998hm,Apanasevich:1998ki}. This could
also explain the differences between the $x_T$ distributions of the
various experiments in fig.~\ref{fig:aurenche_allxs}, since different
experiments run at different energies and are subject to $k_T$ effects
in a different way.
$k_T$ effects give rise to power-like
corrections to the spectrum of order $k_T/p_T$, with possibly very
large coefficients due to to the steepness of the spectrum itself.
The effect of the intrinsic $k_T$ is to smear the $p_T$ distribution,
as shown in fig.~\ref{fig:tung3}.
Inclusion of these effects, however, has a big impact also on the rate
at large $E_T$ (i.e. $x\sim 0.6$).  Due to the large size of the
effects, and to their intrinsic non-perturbative nature (they cannot
be understood from first principles, and need to be described by ad
hoc models), it is hard to trust the theoretical predictions obtained
in this way, and to claim that prompt photons provide a reliable way
of extracting the gluon content of the proton at large $x$. Recent
theoretical improvements, such as the resummation of large-$x_T$
logarithms \cite{Kidonakis:1997gm,Laenen:1998qw,Catani:1998tm,Catani:1999hs},
should help understanding the large-$x$ problem, but more
work is necessary to achieve a satisfactory picture of the data.  
In conclusion, the issue of the large-$x$
behaviour of $f_g(x)$ is still an open problem.

Concerning the possible excess observed by CDF in its highest $\et$
jet data~\cite{cdfjets}, additional input will be available with
the data from the upcoming run of the Tevatron (due to start in the
late 2000), thanks to an increased energy ($\sqrt{S} \to 2$~TeV, 10\%
increase). Should the excess be due to a problem with the gluon
density at large $x$, a discrepancy similar to the one observed at 1.8
TeV will appear at jet $\et$ values larger by 10\%. If the excess is
instead due to really new phenomena, one expects the excess to appear
at the same value of $\et$ as seen in the data at 1.8~TeV. Time will
tell!

\subsection{Cross-section ratios at 630/1800 GeV}
\begin{figure}
\begin{center}
  \includegraphics[width=0.45\textwidth,clip]{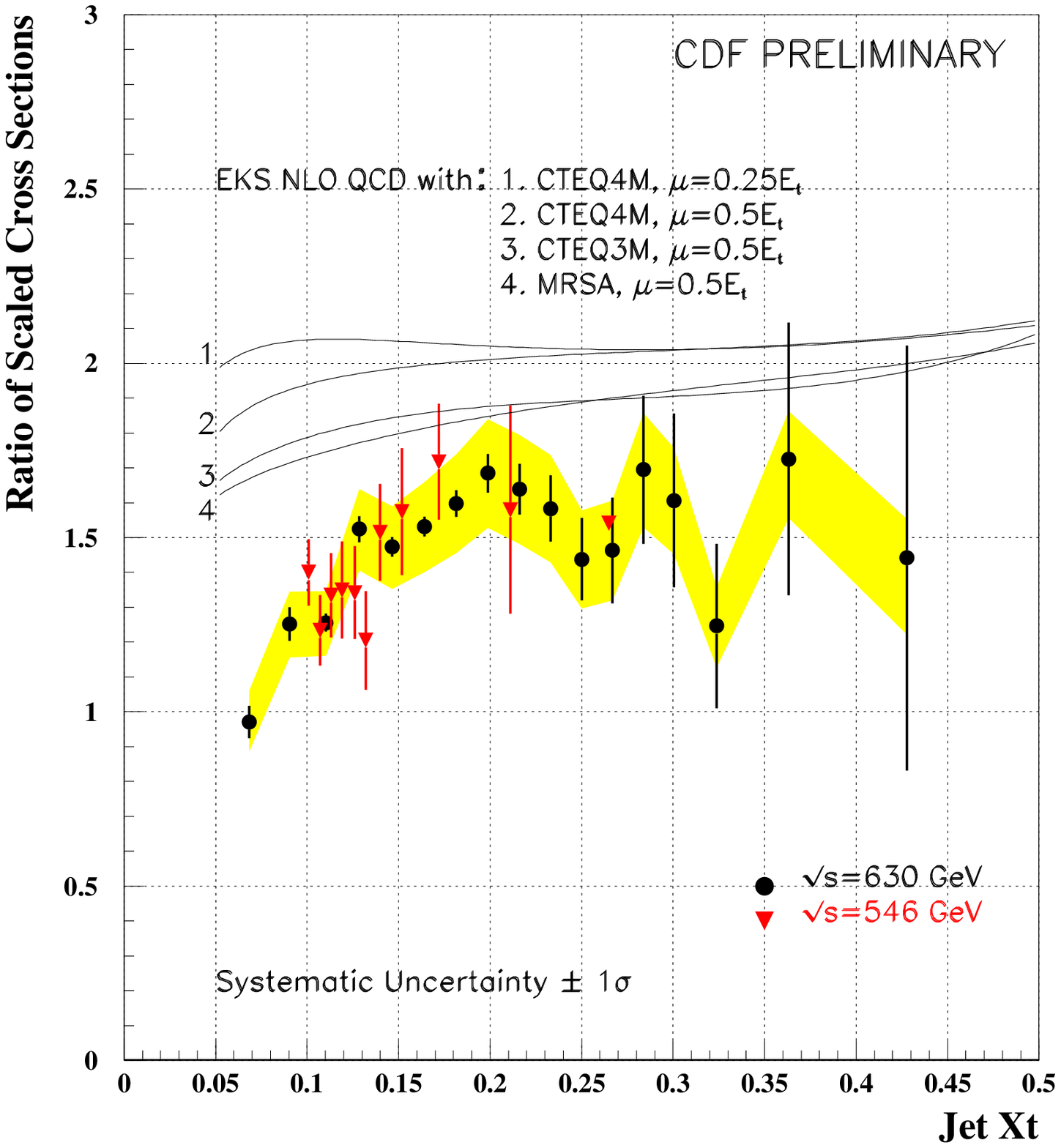}\\
  \includegraphics[width=0.45\textwidth,clip]{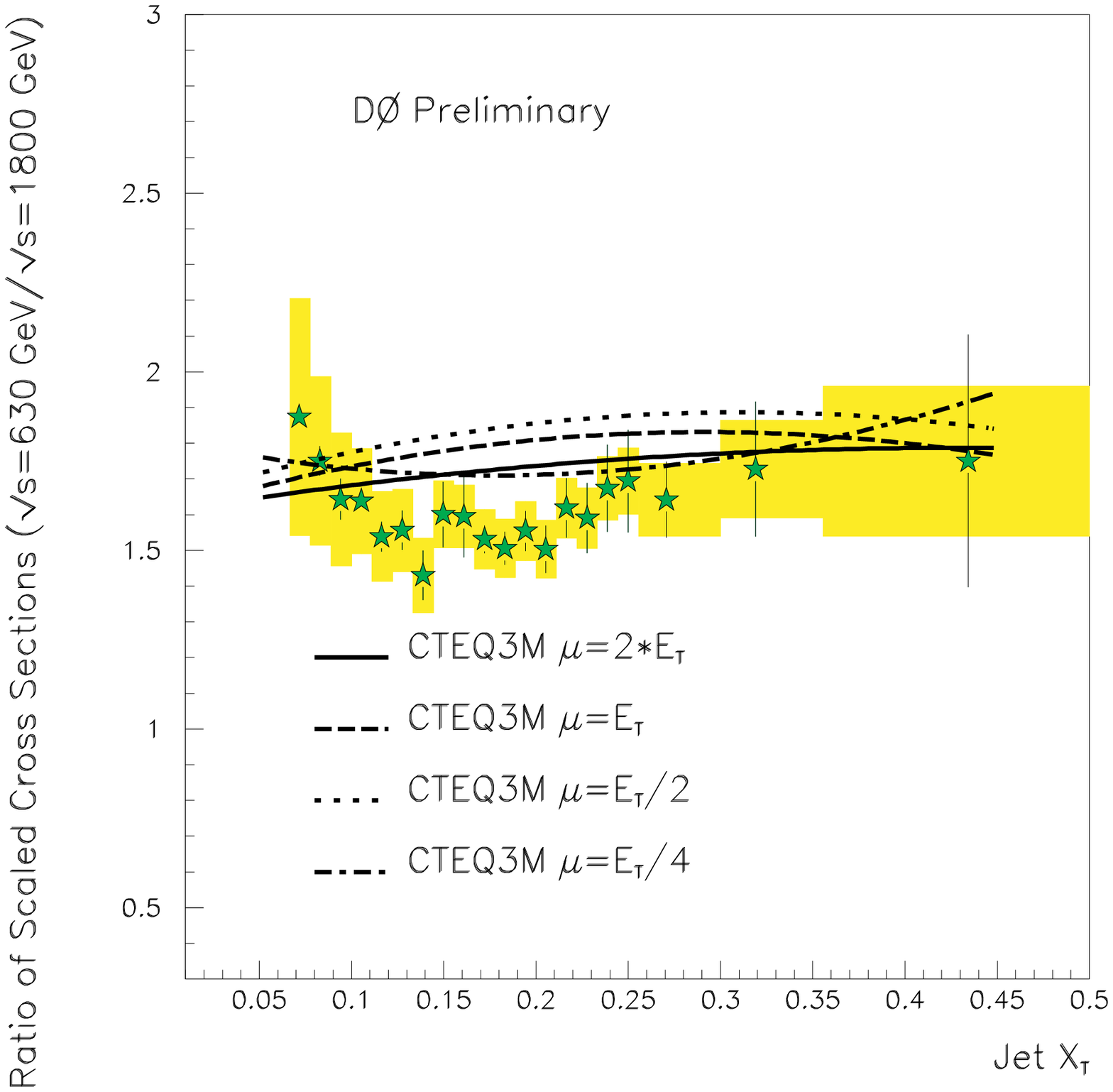}
\end{center}
\vspace*{-0.5cm}
\caption{Cross-section ratios from CDF (upper) and D$\emptyset$ (lower).}
\label{fig:cdf_630to1800}
\end{figure}

It is expected that a large fraction of the theoretical and 
experimental systematics will cancel when taking the ratio:
\be
  R(x_T=\frac{2E_T}{\sqrt{S}}) \=
  \frac{[E_T^3 \; d\sigma/dE_T]_{\sqrt{S}=630}}{[E_T^3 \;  
   d\sigma/dE_T]_{\sqrt{S}=1800}}
\ee
The measurement of $R(x_T)$ can therefore provide a useful additional
tool to explore the physics of high-energy jets.
In the exact scaling limit $R(x_T)=1$. Deviations from 1 arise from
scaling violations in $\as$ and in the parton densities. The NLO
theoretical uncertainty on this ratio is better than 10\%. 
CDF and D$\emptyset$ observe however serious deviations from theory at $x_T\lsim 0.15$
($E_T^{630}\lsim 50$~GeV), as can be seen in fig.~\ref{fig:cdf_630to1800}.
What's more, the pattern of deviations is inconsistent between the
two experiments. I feel that this is a clear indication of the
contamination of the PT results by power-suppressed corrections, as
will be shown in the next section.

\subsection{A pedestrian's evaluation of $x_T$ ratios and power corrections}
To study the theoretical systematics of the $x_T$ distributions, it is
useful to consider a simplified treatment, which however contains all
relevant ingredients. Let us approximate the
inclusive jet rate with the value of the differential
cross-section at $y=0$ for both jets. In this case, at LO, one gets:
\be 
  R(x_T) \= \Sigma(x_T, \; 630~{\rm GeV})/\Sigma(x_T, \; 1800~{\rm GeV})
\ee
with
\be
   \Sigma(x_T,\sqrt{S}) \= \astwo(\mu) \; F^2(x_T,\mu), \quad
    \mu=x_T\sqrt{S}/2 \; .
\ee 
Here 
\be
   F(x)=G(x)+\frac{4}{9}\sum_{q,\bar q} \; \left[ Q(x)+\bar Q(x) \right]
\ee
is the so-called effective structure function for dijet
production~\cite{Halzen:1983rb}. 
Power-suppressed corrections can be included via a factor:
\be
   \Sigma(x_T,\sqrt{S}) \to   \Sigma(x_T,\sqrt{S}) \times
   \left(1 \;+\; \frac{A}{E_T} \right) \;,
\ee
What is the possible origin of $A$, and what is its right order of
magnitude? Here is a partial list of sources: 
\begin{itemize}
\item Energy lost outside the jet cone ($A<0$)
\item Energy from the underlying event inside the jet cone ($A>0$)
\item Intrinsic $k_T$ effects ($A>0$)
\end{itemize}
PT contributions to the energy gain/loss can be evaluated and removed.
However this can be done at LO only, since they are effects of ${\cal
O}(\asthree)$ in PT.  Some energy shifts induced by non-PT effects can
be extracted from the data and corrected for. This is the case, for
example, of the energy deposited in the cone by the Minimum Bias
component of the underlying event.  Correcting for the above effects
may leave us with an $A$ of arbitrary sign, depending on whether one
under- or over-corrects.

In addition, however, there is also a class of non-PT effects which are
out of solid theoretical control, and which cannot be measured in a
direct way. This is the case of parton recombinations with the beam
fragments and with nearby jets.

\begin{figure}
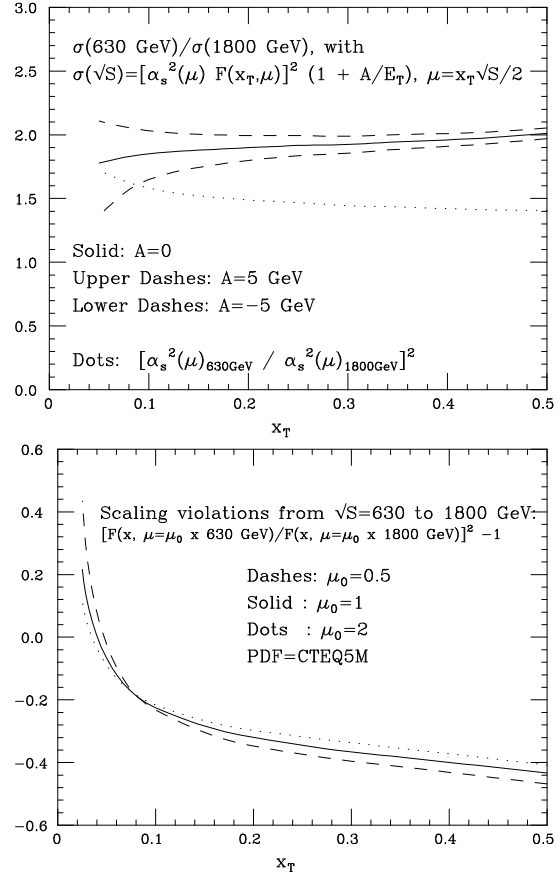

\begin{center}
  \includegraphics[width=0.45\textwidth,clip]{xtrat.eps}\\
  \includegraphics[width=0.45\textwidth,clip]{xtsca.eps}
\end{center}
\vspace*{-0.5cm}
\caption{Upper: effect of power corrections to the scaling violations
in the $x_T$ ratios; the dotted line represents the sole effect due to
the running of $\as(E_T)$. Lower: contribution to $x_T$ scaling
violations due to the parton luminosities.}
\label{fig:xtrat}
\end{figure}

The scale for all these effects is $\Lambda \sim {\cal O}(1~{\rm
  GeV})$. Assuming a $1/E_T^n$ fall-off of the cross-section, one
  gets $A\sim n\Lambda$.  Values of $A\sim5$~GeV should therefore not
  be surprising.  For $A\sim \pm 5$~GeV the effects are large, and can
  be consistent with the deviations observed by CDF and D$\emptyset$, as can be
  seen in fig.~\ref{fig:xtrat}.
Notice that at $x_T\sim 0.05$ all scaling violations are due to the
running of $\as$, since this is an approximate fixed point for the
evolution of the partonic luminosity $F^2(x_T)$.  This is a solid
result, independent of the PDF set chosen, since in this range of
$x_T$ structure functions are known with great accuracy. As a result,
we don't expect that an anomaly in the 630/1800 ratio can be explained
by playing with PDF's.

\begin{figure}
\centerline{
    \includegraphics[width=0.45\textwidth,clip]{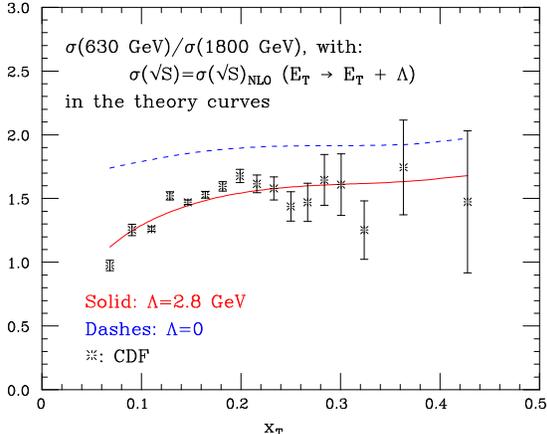}}
\vspace*{-0.5cm}
\caption{Fit of the CDF data using the exact NLO jet 
cross-section (CTEQ3M, $\mu=E_T/2$), assuming  an
$E_T$-independent shift $\Lambda$ in the jet energy.}
\label{fig:xtqfit}
\end{figure}

\begin{figure}
\begin{center}
    \includegraphics[width=0.5\textwidth,clip]{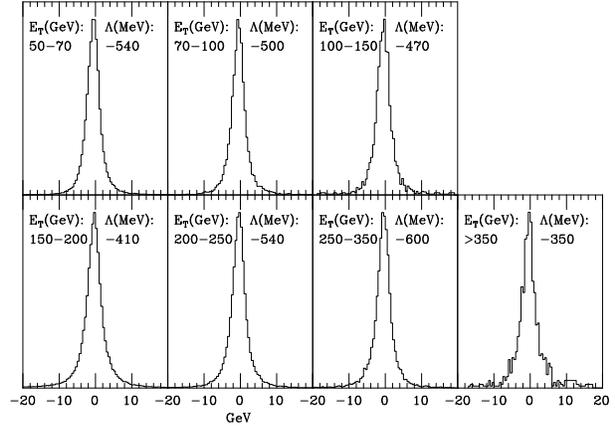}
\vspace*{-0.5cm}
\caption{Hadronisation corrections to the jet energy, for different
ranges of $E_T$. The values of $\Lambda$ indicated 
corresponds to the average of the distributions (in MeV).}
\label{fig:jetnonpt}
\end{center}
\end{figure}

As the results above show, the case is compelling for an
explanation in terms of (acceptably sized) power-like corrections.
For previous studies of power-suppressed effects in the jet
cross-sections and ratios, see e.g. ref.~\cite{Soper:1997xh}, as well
as work in progress by Huston et al.
It is possible to fit the CDF data on $x_T$ ratios using the exact NLO jet 
cross-section (CTEQ3M, $\mu=E_T/2$), assuming a universal and
$E_T$-independent shift in the jet energy. The results of the 
fit are given in fig.~\ref{fig:xtqfit}. 
As can be seen, a shift 
of $-2.8$~GeV relative to the NLO parton-level jet $E_T$
provides a good fit to the CDF data. 
Notice that the effect induced by such a shift is large even at large $x_T$.
Is the size of such a shift acceptable? 
One can estimate the amount of energy lost through the hadronisation
phase using the Herwig MC. Predictions for the quantity
\be
E_{T,jet}^{\rm hadron-level} - E_{T,jet}^{\rm parton-level} 
\ee
are shown in fig.~\ref{fig:jetnonpt}, where I plot the distribution of
the above quantity for jets in several bins of $E_T$.  The corrections
are of the order of 500~MeV, and are remarkably independent of the
value of jet $E_T$, in the range $50<E_T<500$~GeV. This is by itself a
non-obvious result, since in this range the jet compositions in terms
quarks and gluons varies a lot. The 500~MeV are a
non-negligible fraction of what is needed to explain the discrepancies
between CDF/D$\emptyset$ data and NLO theory, confirming the importance of such
phenomena. It would be very interesting to try put on a firmer
theoretical standing the analysis of power corrections to jet spectra
in hadronic collisions, and in particular to explore the relation
between the jet energy correction, and the correction to other
possible observables. Uncovering some universality relation similar to
those found for $e^+e^-$ and DIS observables would open the way to a
new set of interesting measurements in hadronic collisions.
A first study of the effect of power corrections on jet-shape
observables in hadronic collisions was performed by Seymour in 
ref.~\cite{Seymour:1997kj}

\subsection{Conclusions on jet production at the Tevatron}
There is no evidence in my view for departures from QCD in the
 inclusive jet data. Generally good agreement with QCD was also found
 in the studies of multi-jet final states presented by CDF ((1\_595):
 Two-jet Differential Cross Section from CDF; (1\_596): The fully
 corrected dijet invariant mass distribution from CDF) and by D$\emptyset$
 ((1\_163a): The triple differential dijet cross section at D$\emptyset$).

Current discrepancies ($E_T$ spectrum at CDF, $x_T$ ratios 630/1800 at
both CDF and D$\emptyset$) are within theoretical and experimental uncertainties
once proper account is taken of:
\begin{itemize} 
\item  true uncertainties on the extraction of the gluon density
\item  power corrections
\item limitations of the cone algorithm (see
  ref.~\cite{Seymour:1997kj} for a discussion)
\end{itemize} 
In view of this, it is premature in my view to use jets for accurate
measurements, such as the extraction of $\alpha_S(Q^2)$. However,
better use can be made in the future of the large statistics, high
$E_T$ reach, and powerful control of the experimental systematics, if
progress on the theory side can achieve: 
\begin{itemize} 
\item firmer understanding of the intrinsic $k_T$ effects in
fixed-target $\gamma$ production
\item NLL resummations for jet shape variables
\item control (even at the phenomenological level) of the power
corrections
\end{itemize} 
New ideas are needed for observables which can help disentangling the
various components of the theoretical uncertainties.

\section{MULTI-JET PHENOMENA IN $e^+e^-$ and $ep$}
\begin{figure}
\begin{center}
  \includegraphics[width=0.4\textwidth,clip]{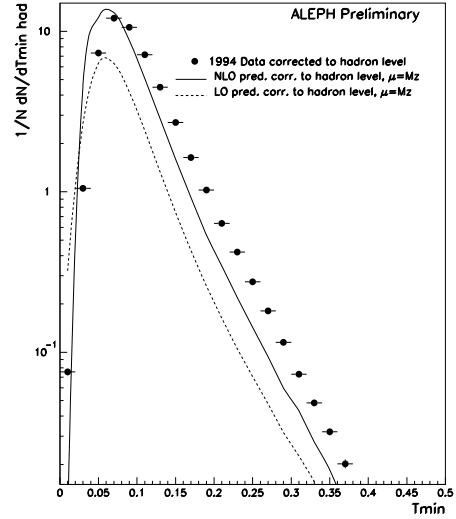} \\
  \includegraphics[width=0.4\textwidth,clip]{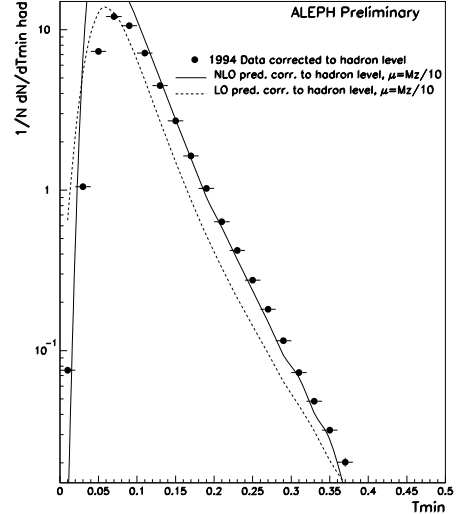}
\end{center}
\vspace*{-0.5cm}
\caption{ALEPH study of the thrust-minor distribution in 4-jet $Z^0$
  decays. Comparison with NLO QCD for different values of $\mu_R$.}
\label{fig:aleph_4jT0}
\end{figure}

Several contributions were submitted with studies of multi-jet
processes. In some cases, new NLO calculations have recently become
available for these observables. Given the large powers of $\as$
involved, however, the scale dependence of the results is still large,
and it is premature to use these data for a better measurement of $\as$.
In all cases, however, the agreement between data and theory
is satisfactory:\\[0.2cm]
(1\_544) ZEUS: Three-jet distributions ($M_{3j}>50$~GeV) 
 in $\gamma$p and LO QCD. \\
(1\_553) ZEUS: NLO tests of high-mass dijet cross-sections in $\gamma$p,
$47<M_{jj}<140$~GeV~\cite{Breitweg:1999wi}. \\
(1\_531) ZEUS: Dijet cross-sections in DIS. \\
(1\_540) ZEUS: Dijet X-sections in $\gamma$p. \\
(1\_386) ALEPH: NLO tests for 4-jet observables in $Z^0$ decays. This
measurement could be important to definitely rule out the existence of
light gluinos, whose presence would affect these
distributions\cite{deGouvea:1997va}.  The comparison of data and
theory (for a recent review, and extensive references, see
\cite{Weinzierl:1999yf}) is shown in fig.~\ref{fig:aleph_4jT0} for the
specific example of the $T_{min}$ (Thrust minor) variable. No good
overall fit to the data can be obtained, even at the price of changing
the renormalisation scale over a wide range. No conclusion on the
issue of a light gluino can however be drawn from these data either.

\section{GAUGE BOSON PRODUCTION in $p\bar p$ and $ep$}
Production of gauge bosons (photons, W and Z) is an extremely useful
tool to probe several aspects of QCD. First of all the clear
experimental signatures allow for very efficient trigger strategies,
and make it possible to probe regions of phase-space were subtle QCD
effects become evident. This is the case, for example, of the $p_T$
distributions of $W$ and $Z$ bosons, which can be probed in hadronic
collisions down to small values of $p_T$, in a domain where
perturbative resummation and intrinsic $k_T$ effects are
dominant~\cite{altarelli}.
Large amounts of theoretical work have been done
recently~\cite{Wptrecent,Corcella:1999gs}. The data (CDF (1\_601) and
D$\emptyset$ (1\_71d)~\cite{Abbott:1999yd}-\cite{Ellison:1999jx}) 
confirm the need of resummation of large
logarithms of $M_{W,Z}/p_T$, as well as the presence of an additional
intrinsic $k_T$, of the order of 2~GeV. we all look forward to 
the quality and statistics of future data from the Tevatron, which will
allow stringent tests of the different theoretical
approaches~\cite{Corcella:1999gs}. 

The presence of a small intrinsic $k_T$ contribution is also advocated
in the case of prompt photon production, similarly to the fixed-target
case discussed above. These results were covered in the following
contributions:\\ 
{ (1\_599) CDF:} Measurement of the Isolated Photon Cross Section~\cite{Kuhlmann:1999xk}.
\\
{ (1\_598) CDF:} Diphoton Production~\cite{Kuhlmann:1999xk}.
\\ 
{ (1\_531) ZEUS:} Prompt Photon Processes in
Photoproduction~\cite{Breitweg:1999su}. 

\section{PRODUCTION OF HEAVY QUARKS}
\subsection{HVQ's in $\gamma\gamma$ collisions at LEP2}
\begin{figure}
\centerline{
\includegraphics[width=0.5\textwidth,clip]{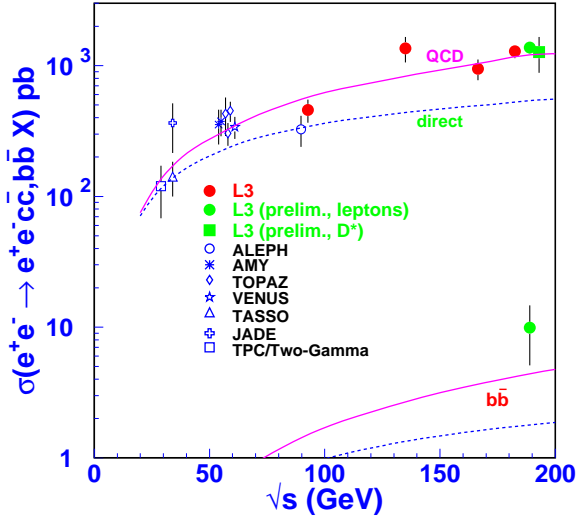}}
\vspace*{-0.5cm}
\caption{L3 measurements of the heavy quark production rates in
$\gamma\gamma$ collisions.}
\label{fig:l3_hvq_gamma}
\end{figure}

Beautiful results have appeared, including the first measurements of
$b\bar b$ production by L3.  Here is a brief summary: \\ 
(1\_265) L3:
$c\bar c $ and $b\bar b$ production in $\gamma\gamma$ at 91-189 GeV
(see fig.~\ref{fig:l3_hvq_gamma}) \\ 
(1\_275)
L3~\cite{Acciarri:1999md}: $D^{*}$ production and $p_T$ spectra in
$\gamma\gamma$ at 183-189 GeV\\ 
(1\_23) OPAL: $D^{*}$ production and
$p_T$ spectra in $\gamma\gamma$ at 183-189 GeV \\ 
All papers share the
same conclusion: the agreement with QCD is very good, provided the
resolved component of the $\gamma$ is included. The accuracy of the
measurements is however not yet sufficient to uniquely disentangle the
gluon density of the photon. 

\subsection{$g\to c\bar c,\; b\bar b$ splitting fractions in $Z^0$ decays}
In addition to providing an interesting playground for studies of PT
QCD~\cite{Mueller:1986zp,Miller:1998ig}, the production of
heavy quark pairs via gluon splitting during a jet evolution provides
also an important contribution to the accurate measurement of
important electroweak properties of heavy quarks, such as
$R_b$~\cite{LEPHF}. Studies of the gluon-splitting fraction have
therefore been an important subject for research by the LEP
collaborations. An important thing to keep in mind however is that these
fractions should not to be used as {\em universal} gluon-splitting
probabilities, as they strongly reflect the spectrum of gluons in
$Z\to$ jets.
Several new results from the latest analyses of
LEP1 data have been shown in Tampere:
\\[0.2cm]
(1\_9, 1\_10) OPAL~\cite{Abbiendi:1999sx}: 
\ba     \langle n_{Z^0}(g\to c\bar c)\rangle \cdot 10^{2} &=&
     3.20(21)(38)
\nn \\
\langle n_{Z^0}(g\to b\bar b)\rangle \cdot 10^{3} &=& 
     2.15(43)(80) \nn
\ea 
(1\_281) L3:
\ba   && \langle n_{Z^0}(g\to c\bar c)\rangle \cdot 10^{2} =\nn \\
  &&     [2.45(35)(45)-3.74(n_{b\bar b}-0.26)] \nn
\ea
(1\_226) DELPHI~\cite{Abreu:1999qh}:
\[ \langle n_{Z^0}(g\to b\bar b)\rangle \cdot 10^{3} = 3.3\pm1.0\pm0.7 
\] 
(1\_184) SLD~\cite{Abe:1999qg}:
\[ \langle n_{Z^0}(g\to b\bar b)\rangle \cdot 10^{3} = 3.07(71)(66)  \]
These numbers are to be compared with the most recent QCD
determinations~\cite{Miller:1998ig} (NLL, $\as=0.120$): 
\[ n_{Z^0}(g\to Q\bar
    Q)\rangle=\begin{array}{ll}
     (m_c=1.2)        &  (m_c=1.5) \\
    2.3\%  & 1.7\%  \\[0.2cm]
     (m_b=4.5)        &   (m_b=4.75) \\
    0.27\%  & 0.24\%  \end{array}
\]
The QCD predictions are on the low side of the experimental results.
One should notice however that the experimental detection efficiencies
are very small, $\calO(\mbox{few \%})$, and therefore require a large
theoretical extrapolation to extract the full rate. In
ref.~\cite{Miller:1998ig} it was pointed out that agreement between
resummed QCD and shower MC's (used for the experimental analyses) is
rather marginal, and it is therefore likely that the true systematics
are larger than quoted.

\subsection{$c,b$ quark production at HERA}
\begin{figure}
\centerline{\includegraphics[width=0.5\textwidth,clip]{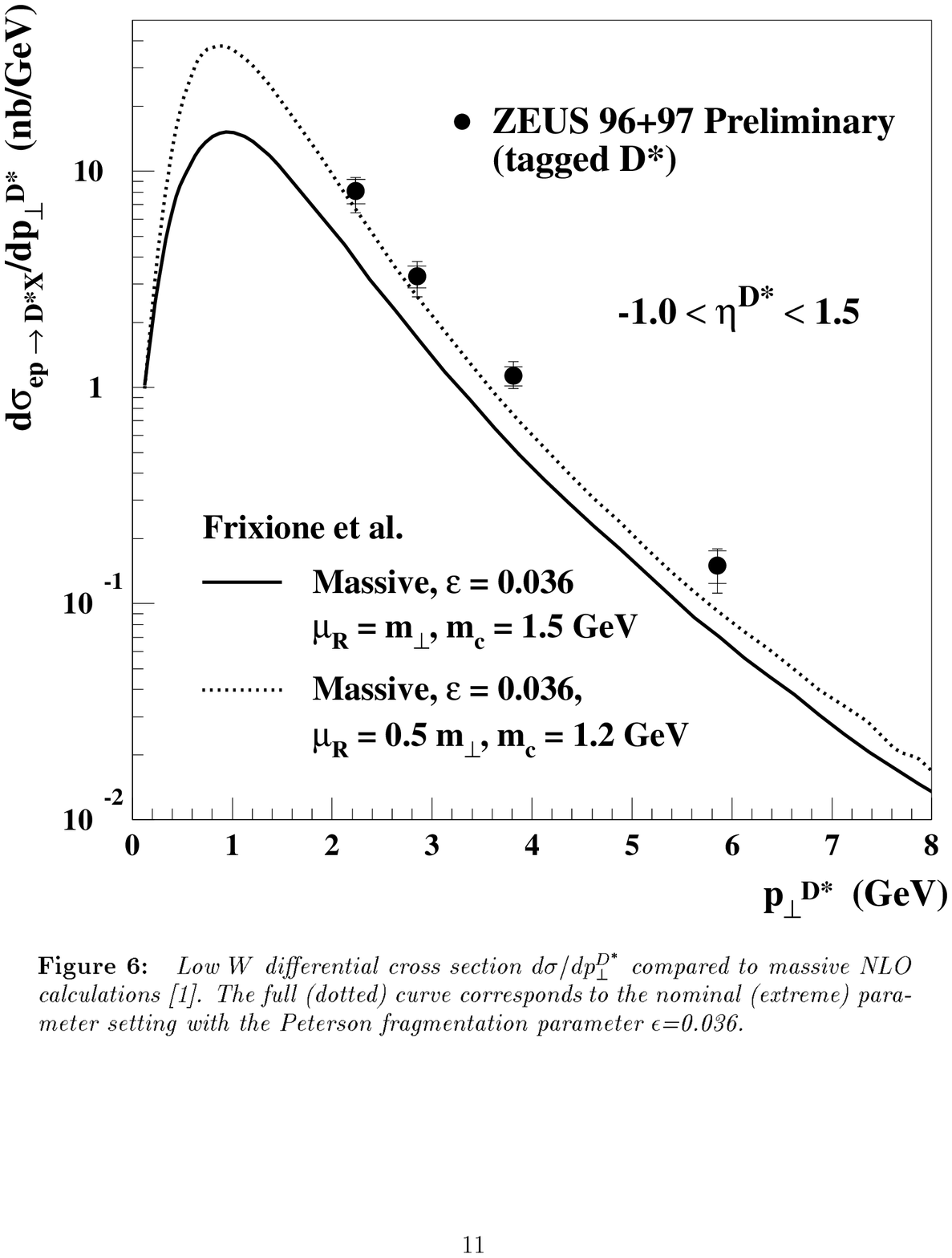}}
\vspace*{-0.5cm}
\caption{$D^{*}$ spectra from ZEUS, compared to NLO QCD calculations
  for different values of the input parameters. $\epsilon$ corresponds
  to the parameter of the Peterson fragmentation function.}
\label{fig:dstarpt}
\end{figure}

Measurements of charm photo-production at HERA have been
available for some time already~\cite{Adloff:1998vb,Breitweg:1998yt}. 
Updates of these measurements have
been presented at this Conference (ZEUS~\cite{Breitweg:1999ad}, 
(1\_525) and (1\_528)),
including studies of $D_s^{\pm}$ production. The agreement between data
and massive NLO QCD calculations~\cite{Frixione:1995dv,Frixione:1995qc}
is generally good, compatibly with the theoretical
uncertainties due to the choice of the charm quark mass, and of the
renormalisation/factorisation scales. H1 finds good agreement for all
kinematical variables considered, while ZEUS (which however has a
looser cut in $Q^2$, and therefore a less strict definition of
photo-production), finds some slight excess in the forward region and
for $p_T>3$~GeV. As an example of the quality of the comparisons,
the ZEUS $p_T$
spectrum for $D^{*}$ mesons is shown in Fig.~\ref{fig:dstarpt}. 
Theoretical estimates including the resummation of large-$p_T$
logarithms, done in a framework 
where the charm is treated as a massless parton, up to mass thresholds
built into the evolution of structure and fragmentation functions,
have recently been performed~\cite{Cacciari:1997du,Binnewies:1998xq}.
Attempts to describe the current data using this approach, however,
would not seem justified in my view in the range of $p_T$ accessible
to the experiments, as corrections of $\calO(m^2/(p_T^2+m^2))$ are not
negligible. By definition, if the massive and the massless calculation
were to differ in the region $p_T<$~few$\times m_c$, one should trust
the massive result.  It will be therefore interesting to see the
results of calculations where the massive cross-sections are matched
at high $\pt$ to the resummed expressions, similarly to what was done
in refs.~\cite{Olness:1997yc,Cacciari:1998it} in the case of
hadro-production.  For a review of the theoretical status, see e.g.
ref.~\cite{Frixione:1999xz}.

Studies have also been presented on bottom photoproduction at
HERA~\cite{Hayes:1999ie}. The results here are more puzzling than in
the case of charm.  Both H1~\cite{Adloff:1999nr} (5\_157v) and ZEUS
(1\_498) tag $b$ events using semileptonic decays. Charm and
fake-lepton backgrounds are separated using the transverse momentum
distribution of the lepton relative to the jet direction. The same
selection and cuts applied to the data are used on samples of MC
events generated using LO matrix elements for the $b\bar b$
photoproduction. The comparison of observed and expected rates is
reported here: 
\\[0.2cm]
(5\_157v) H1 single lepton (LO=Aroma MC):
\[ \sigma^{vis}_{b\bar b}({\rm nb}) = \begin{array}{ll}
 0.93\pm0.08_{stat}{\sst +0.21\atop \sst-0.12}_{syst} & \mbox{H1 Data} \\
 0.19 & \mbox{LO} \end{array}
\]
(5\_157v) H1 dimuons (LO=Aroma MC):
\[ \sigma^{vis}_{b\bar b}({\rm pb}) = \begin{array}{ll}
 55\pm30_{stat}\pm 7_{syst}& \mbox{H1 Data} \\
 17 & \mbox{LO} \end{array}
\]
(1\_498) ZEUS single lepton (LO: Herwig MC):
\[ \sigma^{vis}_{b\bar b}({\rm pb}) = \begin{array}{ll}
 39\pm11_{stat}{\sst +0.23\atop \sst-16}_{syst} & \mbox{ZEUS Data} \\
 10 & \mbox{LO} \end{array}
\]
The results indicate an excess of data relative to the LO QCD
expectations by a factor of approximately 4. In my view it is very
hard to accept this result, given the excellent agreement observed in
the case of charm production. Bottom production at HERA is expected to
be more reliably estimated than charm
production~\cite{Frixione:1995dv}, due to the larger bottom mass. It
will be interesting to see how these studies evolve once more
statistics will have become available. More accurate studies of the
bottom quark distributions will be necessary to validate the MC tools
used by H1 and ZEUS to estimate the detection acceptances and
efficiencies.

\subsection{Bottom quark production at the Tevatron}
\begin{figure}
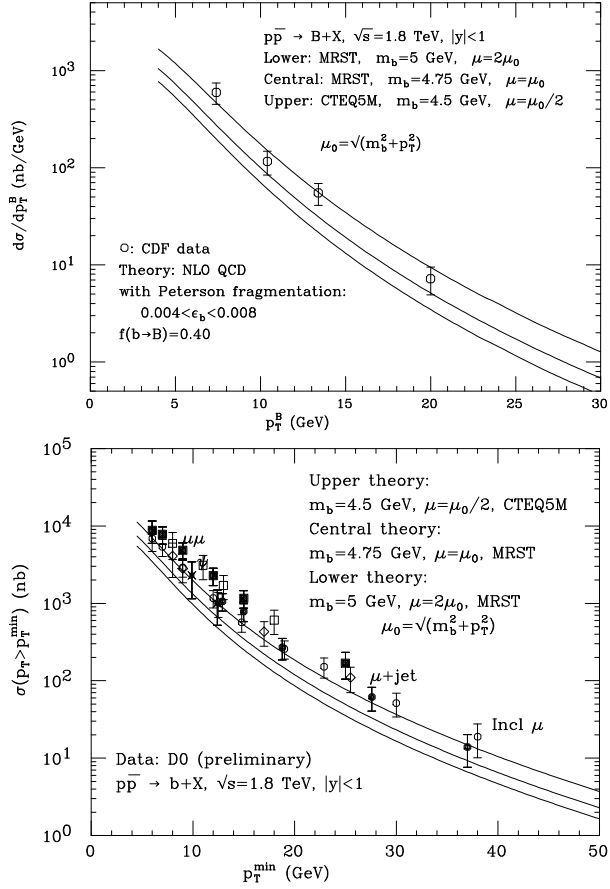

\begin{center}
  \includegraphics[width=0.5\textwidth,clip]{bptmes.eps} \\
  \includegraphics[width=0.5\textwidth,clip]{d0binc.eps}
\end{center}
\vspace*{-0.5cm}
\caption{Left: differential $B$-meson $p_T$ spectra from CDF,
(1\_37)). Right: integrated $b$-quark $p_T$ spectra
from D$\emptyset$~\cite{Abbott:1999se}.}
\label{fig:bptmes}
\end{figure}

The prediction of bottom cross-sections in hadronic collisions is a
sore point for perturbative QCD. NLO calculations have been available
for several years now for the total cross sections~\cite{Nason:1988xz}, for
single-inclusive distributions~\cite{Nason:1989zy} and for
correlations~\cite{Mangano:1992jk}. As pointed out in the original
papers~\cite{Nason:1988xz}, the inclusion of NLO corrections increases the
rates by factors of order 2, and leaves a large scale dependence (of
order 2, and more if renormalisation and factorisation scales are
varied independently). As a result, any comparison with data (for a
recent complete review, see ref.~\cite{Frixione:1997ma}) will at best
be qualitative, and certainly will not provide a compelling test of
the theory. The current comparison with NLO QCD of single-inclusive
rates, as measured by CDF (1\_37) and D$\emptyset$, is summarised in
Fig.~\ref{fig:bptmes} (differential $B$-meson $p_T$ spectra from CDF
and integrated $b$-quark $p_T$ spectra from D$\emptyset$~\cite{Abbott:1999se}).

Within the theoretical uncertainties, the agreement with data is
acceptable. The comparison indicates that smaller values of the
renormalisation and factorisation scales are favoured. Indeed, if one
were to push the scale down to values of the order of
$\sqrt{m_b^2+p_T^2}/4$, the theory curve would exactly overlap the
data.  In spite of the large uncertainty in the prediction of the
absolute rates, the NLO predictions for the shapes of the $b\bar b $
correlations are better defined. Evidence was given in the
past~\cite{Frixione:1997ma}, and confirmed recently
in~\cite{Abbott:1999se}, that NLO QCD provides a good description of
the shape of azimuthal $b\bar b$ correlations. It was shown by
CDF in this conference, (1\_123), that the theory provides also a good
description of the $b\bar b$ rapidity correlations~\cite{Abe:1998ac}. 
All of these observations make therefore rather intriguing the
anomaly observed by D$\emptyset$ in the inclusive forward
production of $b$ quarks~\cite{Abbott:1999wu}. In this paper, D$\emptyset$
reports a factor of 2 excess in the production of forward b's,
relative to what expected by extrapolating the rate measured in the
central rapidity region. Possible mechanisms have been proposed
to increase the expected rates for forward production of $B$ mesons
(e.g. a harder non-perturbative fragmentation function
\cite{Mangano:1997ri}), but none of them can explain the large effect
observed by D$\emptyset$.

\begin{figure}
\centerline{
  \includegraphics[width=0.5\textwidth,clip]{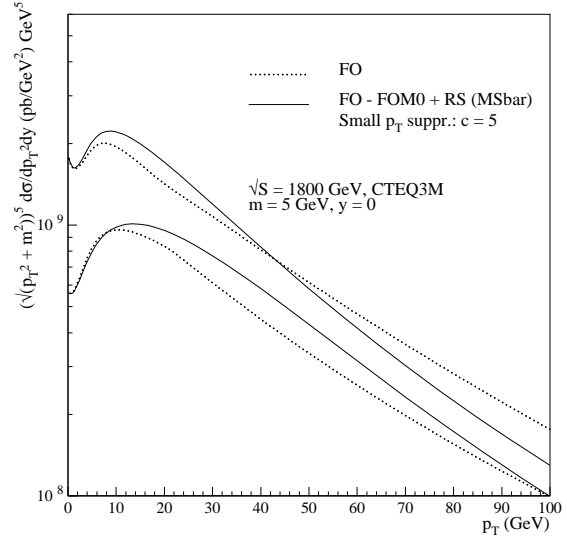}}
\vspace*{-0.5cm}
\caption{Improvement in the scale dependence of the bottom quark $p_T$
  spectrum after inclusion of large-$p_T$ resummation contributions
  (Cacciari et al., \cite{Cacciari:1998it}. The bands correspond to
  changes of $\mu$ in the range $m_T/2<\mu<2m_T$, with
  $m_T^2=m^2+p_T^2$ (dotted lines: NLO; solid lines: NLO+resummed).}
\label{fig:cacciari_gn}
\end{figure}

Progress in theory took place in the recent past. This includes
studies of resummation of the large logarithms of $p_T/m_b$ which
appear at any order of PT~\cite{Olness:1997yc,Cacciari:1998it}.  The
accuracy of  these calculations is now full $\calO(\asthree)$ plus
the resummation of NL logarithms arising in the fragmentation function
of the heavy quark~\cite{Mele:1991cw}.  The results (see
fig.~\ref{fig:cacciari_gn}) show an improved scale dependence at large
$p_T$, as expected, and an increase in rate in the region of
$10\gsim\pt\lsim 40$~GeV, where most of the data from CDF and D$\emptyset$
are sitting. Unfortunately the resummation of this class of logarithms
does not provide reliable information in the region of $\pt\lsim
20$~GeV, since in this region mass corrections have been shown to be
large~\cite{Cacciari:1998it}. The large $K$-factor shown in
fig.~\ref{fig:cacciari_gn} at small $p_T$ cannot be used, therefore,
to improve the agreement between data and theory.

On the front of the experimental inputs, I should recall here the
recent accurate experimental studies of $B$-meson spectra at LEP and
SLD~\cite{Abe:1999fi} (1\_182). These will provide the necessary
ingredients for accurate extractions of the non-perturbative component
of the fragmentation functions, as discussed recently for example in
ref.~\cite{Nason:1999zj}.

\subsection{Top quark production at the Tevatron}
\begin{figure}
\begin{center}
\includegraphics[width=0.5\textwidth,clip]{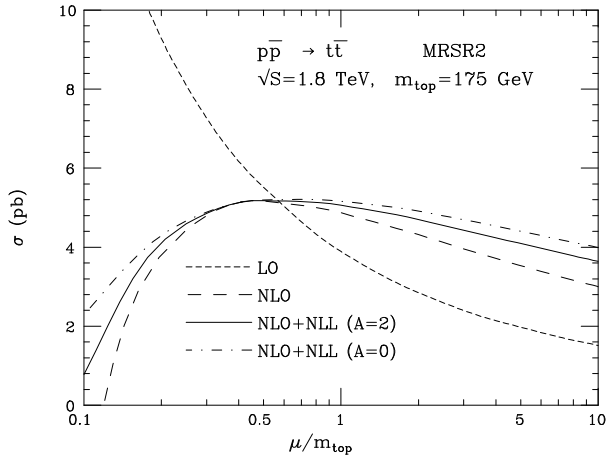}
\end{center}
\vspace*{-0.5cm}
\caption{Scale dependence of $\sigma_{t\bar t}$ at the Tevatron (1.8
  TeV), for various degrees of accuracy in the QCD calculation.}
\label{fig:tscale}
\end{figure}

Theoretical predictions for $t\bar t $ production at the
Tevatron are expected to be rather robust, given the large value of
the top mass and the correspondingly small value of the coupling,
$\as(m_{top})$, appearing in the QCD perturbative expansion.
The next-to-leading-log (NLL) resummation of Sudakov threshold effects
has been carried out in the past year~\cite{Kidonakis:1997gm,Bonciani:1998vc}
Results indicate a good reduction in scale uncertainty, to the level of
$\pm 5\%$, as shown in Fig.~\ref{fig:tscale}~\cite{Bonciani:1998vc}

In addition to the scale-variation uncertainty, a $\pm 7\%$ variation
in the theoretical predictions is present due to the choice of PDF's
($\sigma$'s in pb):
\[
\begin{array}{llll}
  \mbox{PDF} &  \mu=m_{t}/2  &  \mu=m_{t}  &  \mu=2m_{t}  \\
  \mbox{MRST} &   5.04    &   4.92    &   4.57    \\
  \mbox{MRST} g \uparrow  &   5.22    &   5.09    &   4.72    \\
  \mbox{MRST} g \downarrow  &   4.90    &   4.79    &   4.45    \\
  \mbox{MRST}  \as\downarrow  & 4.84
                             & {4.74}  &   4.42    \\
  \mbox{MRST}  \as\uparrow   &   5.20    &   5.07    &   4.68   \\
  \mbox{CTEQ5M} & 5.41    &   5.30    &   4.91    \\
  \mbox{CTEQ5HJ}&   5.61    & {5.50} &   5.10    
\end{array}
\]
(MRST: \cite{Martin:1998sq}; CTEQ5:~\cite{Lai:1999wy}; NLO+NLL results
from~\cite{Bonciani:1998vc}, using the prescription for the inverse Mellin
transform introduced in~\cite{Catani:1996yz})

At this Conference a new determination of the $t\bar t $ cross-section
measured by CDF was presented by Ptohos~(5\_455). The new value is
approximately 1 standard deviation lower than the previous one, and in
much better agreement with the QCD predictions. The overall
cross-section averages from CDF and D$\emptyset$ (in this last case rescaled to
$m_{top}=175$~GeV) are shown in Table~\ref{tab:sigmatop}, and compared
to various theoretical results appeared in the literature. Now that
the CDF number has come down a bit, the average of the experimental
determinations ($5.9\pm 1.3$~pb at 175~GeV) is within less than one standard
deviation from the QCD NLO+NLL resummed result of ($5.0\pm 0.6$~pb)
(\cite{Bonciani:1998vc}, with scale and PDF uncertainties added
linearly). It is interesting to
notice that both CDF and D$\emptyset$~\cite{Abbott:1998nn} report significantly
lower values for $\sigma(t\bar t)$ in the single-lepton plus jets
channels than in the all-jet or dilepton ones. These lower values are
in closer agreement with QCD than the overall average. It is clearly
premature to draw any conclusion on this small discrepancy between the
determinations obtained using the various channels. 
\begin{table}
\label{tab:sigmatop}
\begin{center}
\caption{$t\bar t$ cross-sections, in pb, for $m_{top}=175$~GeV. Upper
  rows: CDF:
  (5\_455); D$\emptyset$: \cite{Abbott:1998nn}. Lower rows:
  BCMN: \cite{Catani:1996yz}; BC:
  \cite{Berger:1998gz}; K: \cite{Kidonakis:1998qe}}  
\vspace*{0.2cm}
\begin{tabular}{ccc}
\hline
  CDF &   D$\emptyset$ & \\ 
$6.5\pm 1.5$ &    
$5.4\pm 1.5$ & \\ \hline\hline
  BCMN &   BC &   K \\
$5.0\pm 0.6$ &
$5.57{\sst +0.07\atop \sst -0.42}$ & 
7
\\ \hline
\end{tabular}
\end{center}
\end{table}
Several studies of kinematical properties of top final states have
been presented by CDF~\cite{Koehn:1999tz} and D$\emptyset$. All results are in
good agreement with the predictions from NLO QCD~\cite{Frixione:1995fj}.

\section{CONCLUSIONS}
After over 25 years since its discovery, QCD is still a very rich an
exciting field, with progress both in the experimental techniques and
in the theoretical understanding. Measurements are becoming more and
more sophisticated, and the challenge for theorists is becoming harder
and harder.  The accuracy in the extraction of $\as$ is reaching its
limits. New theoretical developments will be necessary to take full
advantage of the future $ep$ and $p \bar p$ (as well as LHC) data.

A consistent phenomenological picture of the impact of the
hadronisation phase on the structure of final states is emerging.
Tests in $e^+e^-$ collisions are becoming very compelling, and the
universality of the description of power corrections has been tested
even in $ep$ collisions, to a level of accuracy which is consistent
with the current expectations. At this time, however, the
uncertainties in the determination of $\as$ performed using the
analytic description of hadronisation effects are not smaller than
those found using the Monte Carlo modelling. The largest source of
theoretical uncertainties in the extraction of $\as$ still remains the
renormalisation-scale dependence of the results. New approaches, such
as the experimentally optimisation of the scale pursued by DELPHI,
will have to await more solid theoretical justification before their
potential can be fully exploited.

Application of these ideas to hadronic collisions will require more
work. The new frontier is the evaluation of NNLO cross-sections and
NLL resummations, and applications to the study of jet shapes. A
recent calculation of 3-jet production at
$\calO(\asfour)$~\cite{Kilgore:1997sq} could be
used for a first evaluation of jet shapes at NLO.
 Attention should go to the use of appropriate jet
algorithms, and to the identification of appropriate observables.
Extraction of the gluon density of the proton from photon and jet data
has also reached the limit of theoretical accuracy. Progress on the
above points will be necessary before further improvements can be
achieved.


\begin{thebibliography}{99}
\bibitem{ALEPH}
 Aleph Coll., http://alephwww.cern.ch/ \\ ALPUB/conf/conf.html
\bibitem{DELPHI}
 DELPHI Coll.,
 {\tt http://delphiwww.cern.ch/\~{}pubxx/\\
  delwww/www/delsec/conferences/tampere99/}
\bibitem{L3}
 L3 Coll., {\tt http://l3www.cern.ch/conferences/EPS99/}
\bibitem{OPAL}
 OPAL Coll., {\tt http://www.cern.ch/Opal/\\ pubs/eps99\_sub.html}
\bibitem{SLD}
 SLD Coll., {\tt http://www-pnp.physics.ox.ac.uk/ \\ \~{}burrows/tampere/}
\bibitem{ZEUS}
 ZEUS Coll., {\tt http://zedy00.desy.de/conferences99/}
\bibitem{H1}
 H1 Coll., {\tt 
 http://www-h1.desy.de/h1/www/ \\ publications/conf/list.tampere99.html}
\bibitem{D0}
 D$\emptyset$ Coll., 
 {\tt http://www-d0.fnal.gov/\\ \~{}ellison/eps99/eps99.html}
\bibitem{Webber}
  B. Webber, rapporteur talk at the 1999 Lepton-Photon Symposium,
  SLAC, 1999.
\bibitem{Draggiotis:1998gr}
  P.~Draggiotis, R.H.~Kleiss and C.G.~Papadopoulos,
  Phys. Lett. {\bf B439}, 157 (1998)
  hep-ph/9807207.
  F.~Caravaglios, M.L.~Mangano, M.~Moretti and R.~Pittau,
  Nucl. Phys. {\bf B539}, 215 (1999)
  hep-ph/9807570.

\bibitem{Bern:1998sc}
  Z.~Bern, V.~Del Duca and C.R.~Schmidt,
  Phys.\ Lett.\ {\bf B445}, 168 (1998)
  hep-ph/9810409.
  Z.~Bern, V.~Del Duca, W.B.~Kilgore and C.R.~Schmidt,
  Phys.\ Rev.\ {\bf D60}, 116001 (1999)
  hep-ph/9903516.
  D.A.~Kosower and P.~Uwer,
  hep-ph/9903515.
  J.M.~Campbell and E.W.~Glover,
  Nucl. Phys. {\bf B527}, 264 (1998)
  hep-ph/9710255.
  S.~Catani,
  Phys.\ Lett.\ {\bf B427}, 161 (1998)
  hep-ph/9802439.
    S.~Catani and M.~Grazzini,
    Phys. Lett. {\bf B446}, 143 (1999)
    hep-ph/9810389.
    S.~Catani and M.~Grazzini,
    hep-ph/9908523.
\bibitem{Chetyrkin:1999bx}
  K.G.~Chetyrkin, R.~Harlander and J.H.~Kuehn,
  hep-ph/9910345.
\bibitem{Kidonakis:1997gm}
  N.~Kidonakis and G.~Sterman,
  Nucl. Phys. {\bf B505}, 321 (1997)
  hep-ph/9705234.
\bibitem{Bonciani:1998vc}
  R.~Bonciani, S.~Catani, M.L.~Mangano and P.~Nason,
  Nucl. Phys. {\bf B529}, 424 (1998)
  hep-ph/9801375.
\bibitem{Kidonakis:1998bk}
  N.~Kidonakis, G.~Oderda and G.~Sterman,
  Nucl. Phys. {\bf B525}, 299 (1998)
  hep-ph/9801268.
\bibitem{Laenen:1998qw}
  E.~Laenen, G.~Oderda and G.~Sterman,
  Phys. Lett. {\bf B438}, 173 (1998)
  hep-ph/9806467.
\bibitem{Catani:1998tm}
  S.~Catani, M.L.~Mangano and P.~Nason,
  JHEP {\bf 07}, 024 (1998)
  hep-ph/9806484.
\bibitem{Catani:1999hs}
  S.~Catani, M.L.~Mangano, P.~Nason, C.~Oleari and W.~Vogelsang,
  JHEP {\bf 03}, 025 (1999)
  hep-ph/9903436.
\bibitem{Kidonakis:1999xm}
  N.~Kidonakis,
  hep-ph/9910240.
\bibitem{Sterman:1999gz}
  G.~Sterman and W.~Vogelsang,
  hep-ph/9910371.
\bibitem{Bernreuther:1997jn}
  W.~Bernreuther, A.~Brandenburg and P.~Uwer,
  Phys.\ Rev.\ Lett.\ {\bf 79}, 189 (1997)
  hep-ph/9703305.
\bibitem{Rodrigo:1997gy}
  G.~Rodrigo, A.~Santamaria and M.~Bilenkii,
  Phys.\ Rev.\ Lett.\ {\bf 79}, 193 (1997)
  hep-ph/9703358; 
  Nucl.\ Phys.\ {\bf B554}, 257 (1999)
  hep-ph/9905276.
\bibitem{Oleari:1997az}
  C.~Oleari,
  hep-ph/9802431.
  P.~Nason and C.~Oleari,
  Nucl.\ Phys.\ {\bf B521}, 237 (1998)
  hep-ph/9709360.
\bibitem{Abreu:1997ey}
  P.~Abreu {\it et al.} [DELPHI Coll.],
  Phys.\ Lett.\ {\bf B418}, 430 (1998).
\bibitem{Brandenburg:1999nb}
  A.~Brandenburg, P.N.~Burrows, D.~Muller, N.~Oishi and P.~Uwer,
  hep-ph/9905495.
\bibitem{Melnikov:1998ug}
        K.~Melnikov and A.~Yelkhovsky,
        Phys.\ Rev.\ {\bf D59}, 114009 (1999)
        hep-ph/9805270.
        A.H.~Hoang,
        hep-ph/9905550.
        M.~Beneke and A.~Signer,
        hep-ph/9906475.
\bibitem{Palla:1999st}
  F.~Palla,
  hep-ex/9910044.
\bibitem{Khoze:1997dn}
  V.A.~Khoze and W.~Ochs,
  Int.\ J.\ Mod.\ Phys.\ {\bf A12}, 2949 (1997)
  hep-ph/9701421.
  Y.L.~Dokshitzer, V.A.~Khoze and S.I.~Troian,
  J.\ Phys.\ {\bf G17}, 1602 (1991).
\bibitem{Muller:1999cy}
  D.~Muller {\it et al.}, [SLD and DELPHI Collab.]
  SLAC-PUB-8257, {\it to appear in these proceedings.}
\bibitem{Ballestrero:1992ed}
  A.~Ballestrero, E.~Maina and S.~Moretti,
  Phys.\ Lett.\ {\bf B294}, 425 (1992).
  Nucl.\ Phys.\ {\bf B415}, 265 (1994)
  hep-ph/9212246.
\bibitem{Abbiendi:1999fs}
  G.~Abbiendi {\it et al.} [OPAL Coll.],
  hep-ex/9904013.
\bibitem{Abe:1999pz}
  K.~Abe {\it et al.} [SLD Coll.],
  Phys.\ Rev.\ {\bf D60}, 092002 (1999)
  hep-ex/9903004; 
  hep-ex/9908027.
\bibitem{Abe:1999qc}
  K.~Abe {\it et al.} [SLD Coll.],
  hep-ex/9908031.
\bibitem{Dokshitzer:1991wu}
  Y.L.~Dokshitzer, V.A.~Khoze, A.H.~Mueller and S.I.~Troian,
 ``Basics of perturbative QCD,''
 {\it  Gif-sur-Yvette, France: Ed. Frontieres (1991)}.
\bibitem{Abbott:1999cu}
  B.~Abbott {\it et al.} [D$\emptyset$ Coll.],
  hep-ex/9908017.
\bibitem{Gary:1999bp}
  J.W.~Gary,
  hep-ex/9909024.
\bibitem{Abreu:1999rs}
  P.~Abreu {\it et al.} [DELPHI Coll.],
  Phys.\ Lett.\ {\bf B449}, 383 (1999)
  hep-ex/9903073.
\bibitem{Abbiendi:1999pi}
  G.~Abbiendi {\it et al.} [OPAL Coll.],
  hep-ex/9903027.
\bibitem{Ellis:1981wv}
  R.K.~Ellis, D.A.~Ross and A.E.~Terrano,
  Nucl.\ Phys.\ {\bf B178}, 421 (1981).
  Phys.\ Rev.\ Lett.\ {\bf 45}, 1226 (1980).
\bibitem{Kunszt:1989km}
  Z.~Kunszt, P.~Nason, G.~Marchesini and B.R.~Webber,
  ``QCD At Lep,''
  {\it Proceedings of the 1989 LEP Physics Workshop, Geneva,
                  Swizterland, Feb 20, 1989}.
\bibitem{Catani:1997vz}
  S.~Catani and M.H.~Seymour,
  Nucl.\ Phys.\ {\bf B485}, 291 (1997)
  hep-ph/9605323; Erratum-ibid.B510:503-504,1997.
\bibitem{Catani:1993ua}
  S.~Catani, L.~Trentadue, G.~Turnock and B.R.~Webber,
  Nucl.\ Phys.\ {\bf B407}, 3 (1993).
\bibitem{Manohar:1995kq}
  A.V.~Manohar and M.B.~Wise,
  Phys.\ Lett.\ {\bf B344}, 407 (1995)
  hep-ph/9406392.
  Y.L.~Dokshitzer and B.R.~Webber,
  Phys.\ Lett.\ {\bf B352}, 451 (1995)
  hep-ph/9504219.
  R.~Akhoury and V.I.~Zakharov,
  Phys.\ Lett.\ {\bf B357}, 646 (1995)
  hep-ph/9504248.
  G.P.~Korchemsky and G.~Sterman,
  hep-ph/9505391.
\bibitem{Dokshitzer:1996qm}
  Y.L.~Dokshitzer, G.~Marchesini and B.R.~Webber,
  Nucl.\ Phys.\ {\bf B469}, 93 (1996)
  hep-ph/9512336.
\bibitem{Abreu:1996mk}
  P.~Abreu {\it et al.} [DELPHI Coll.],
  Z.\ Phys.\ {\bf C73}, 229 (1997).
\bibitem{Adloff:1997gq}
  C.~Adloff {\it et al.} [H1 Coll.],
  Phys.\ Lett.\ {\bf B406}, 256 (1997)
  hep-ex/9706002.
\bibitem{Biebel:1999zt}
  O.~Biebel, P.A.~Movilla Fernandez and S.~Bethke [JADE Coll.],
  Phys.\ Lett.\ {\bf B459}, 326 (1999)
  hep-ex/9903009.
\bibitem{Dokshitzer:1997iz}
  Y.L.~Dokshitzer, A.~Lucenti, G.~Marchesini and G.P.~Salam,
  Nucl.\ Phys.\ {\bf B511}, 396 (1998)
  hep-ph/9707532.
\bibitem{Korchemsky:1995is}
  G.P.~Korchemsky and G.~Sterman,
  Nucl.\ Phys.\ {\bf B437}, 415 (1995)
  hep-ph/9411211.
  Y.L.~Dokshitzer and B.R.~Webber,
  Phys.\ Lett.\ {\bf B404}, 321 (1997)
  hep-ph/9704298.
\bibitem{Dokshitzer:1998qp}
  Y.L.~Dokshitzer, G.~Marchesini and G.P.~Salam,
  Eur.\ Phys.\ J.\ direct {\bf C3}, 1 (1999)
  hep-ph/9812487.
  Y.L.~Dokshitzer, A.~Lucenti, G.~Marchesini and G.P.~Salam,
  JHEP {\bf 05}, 003 (1998)
  hep-ph/9802381.
\bibitem{Nason:1995hd}
  P.~Nason and M.H.~Seymour,
  Nucl.\ Phys.\ {\bf B454}, 291 (1995)
  hep-ph/9506317.
\bibitem{Carli:1999eg}
  T.~Carli,  [H1 and ZEUS Coll.]
  hep-ph/9910360.
\bibitem{Mirkes:1996ks}
  E.~Mirkes and D.~Zeppenfeld,
  Phys.\ Lett.\ {\bf B380}, 205 (1996)
  hep-ph/9511448.
  E.~Mirkes,
  hep-ph/9711224.
  D.~Graudenz,
  hep-ph/9710244.
  B.~Potter,
  Comput.\ Phys.\ Commun.\ {\bf 119}, 45 (1999)
  hep-ph/9806437.
\bibitem{Duprel:1999wz}
  C.~Duprel, T.~Hadig, N.~Kauer and M.~Wobisch,
  hep-ph/9910448.
\bibitem{MovillaFernandez:1999yn}
  P.A.~Movilla Fernandez, O.~Biebel and S.~Bethke,
  hep-ex/9906033.
\bibitem{Ohnishi:1994vp}
  Y.~Ohnishi and H.~Masuda,
  SLAC-PUB-6560 (1994).
\bibitem{DelphiOSV}
  P. Abreu et al. [DELPHI Coll.],
  CERN-EP/99-133.
\bibitem{Bethke:1998ja}
  S.~Bethke,
  hep-ex/9812026.
\bibitem{Abbott:1999gj}
  B.~Abbott {\it et al.} [D$\emptyset$ Coll.],
  hep-ex/9907059.
\bibitem{Forshaw:1999iv}
  J.R.~Forshaw and M.H.~Seymour,
  JHEP {\bf 09}, 009 (1999)
  hep-ph/9908307.
\bibitem{Safonov:1999dr}
  A.N.~Safonov [CDF Coll.], FERMILAB-CONF-99-307-C.
\bibitem{cdfjets}
  F. Abe et al., CDF Collab., Phys.\ Rev.\ Lett.\ {\bf 77}, 438 (1996).
\bibitem{Abbott:1998ya}
  B.~Abbott {\it et al.} [D$\emptyset$ Coll.],
  Phys.\ Rev.\ Lett.\ {\bf 82}, 2451 (1999)
  hep-ex/9807018.
\bibitem{jetsnlo}
    F.\ Aversa, P.\ Chiappetta, M.\ Greco and J.Ph. Guillet,
    Nucl.\ Phys.\ {\bf B327}, 105 (1989);
    S.D.~Ellis, Z.~Kunszt and D.E.~Soper,
    Phys.\ Rev.\ Lett.\ {\bf 64}, 2121 (1990);\\
    W.T.~Giele, E.W.~Glover and D.A.~Kosower,
    Nucl.\ Phys.\ {\bf B403}, 633 (1993)
    hep-ph/9302225.
\bibitem{Gallas:1999yp}
  E.J.~Gallas [D$\emptyset$ and CDF Coll],
  {\it to appear in these Proceedings}.
\bibitem{Martin:1998sq}
  A.D.~Martin, R.G.~Roberts, W.J.~Stirling and R.S.~Thorne,
  Eur. Phys. J. {\bf C4}, 463 (1998)
  hep-ph/9803445.
\bibitem{Lai:1999wy}
  H.L.~Lai {\it et al.}
  [CTEQ Collab.],
  hep-ph/9903282.
\bibitem{Aurenche:1998gv}
  P.~Aurenche, M.~Fontannaz, J.P.~Guillet, B.~Kniehl, E.~Pilon and M.~Werlen,
  Eur.\ Phys.\ J.\ {\bf C9}, 107 (1999) hep-ph/9811382.
\bibitem{Apanasevich:1998hm}
  L.~Apanasevich {\it et al.} [E706 Coll.],
  Phys.\ Rev.\ Lett.\ {\bf 81}, 2642 (1998)
  hep-ex/9711017.
\bibitem{Apanasevich:1998ki}
  L.~Apanasevich {\it et al.},
  Phys.\ Rev.\ {\bf D59}, 074007 (1999)
  hep-ph/9808467.
\bibitem{Halzen:1983rb}
  F.~Halzen and P.~Hoyer,
  Phys.\ Lett.\ {\bf 130B}, 326 (1983).
  B.L.~Combridge and C.J.~Maxwell,
  Nucl.\ Phys.\ {\bf B239}, 429 (1984).
\bibitem{Soper:1997xh}
  D.E.~Soper,
  hep-ph/9706320.
\bibitem{Seymour:1997kj}
  M.H.~Seymour,
  Nucl.\ Phys.\ {\bf B513}, 269 (1998)
  hep-ph/9707338.
\bibitem{Breitweg:1999wi}
  J.~Breitweg {\it et al.} [ZEUS Coll.],
  hep-ex/9905046.
\bibitem{deGouvea:1997va}
  A.~de Gouvea and H.~Murayama,
  Phys.\ Lett.\ {\bf B400}, 117 (1997)
  hep-ph/9606449.
  Z.~Nagy and Z.~Trocsanyi,
  hep-ph/9708343.
\bibitem{Weinzierl:1999yf}
  S.~Weinzierl and D.A.~Kosower,
  Phys.\ Rev.\ {\bf D60}, 054028 (1999)
  hep-ph/9901277.
\bibitem{altarelli}
 G. Altarelli, R.K. Ellis and G. Martinelli, Nucl.\  Phys.\ B143 (1978) 521;
 Nucl.\ Phys.\ B146 (1978) 544 (erratum); Nucl.\ Phys.\ B157 (1979) 461.
 Yu.L. Dokshitzer, D.I. Dyakonov and S.I. Troyan, Phys.\ Rep.\ 58 (1980) 269.
 J. Collins, D. Soper and G. Sterman, Nucl.\ Phys.\ B250 (1985) 199.
 C.T.H. Davies, W.J. Stirling and B.R. Webber, Nucl.\ Phys.\ B256 (1985) 413.
 P.B. Arnold and R. Kauffman, Nucl.\ Phys.\ B349 (1991) 381.
\bibitem{Wptrecent}
 G.A. Ladinsky and C.P. Yuan, Phys.\ Rev.\ D50 (1994) 4239.
 R.K. Ellis, D.A. Ross and S. Veseli, Nucl.\ Phys.\ B503 (1997) 309.
 R.K. Ellis and S. Veseli, Nucl.\ Phys.\ B511 (1998) 649.
 S. Frixione, P. Nason and G. Ridolfi, Nucl.\ Phys.\ B542 (1999) 311.
 A. Kulesza and W.J. Stirling, DTP-99-02, hep-ph/9902234.
 G. Miu and T. Sj\"ostrand, Phys.\ Lett.\ B449 (1999) 313.
 S. Mrenna, UCD-99-13, hep-ph/9902471.
\bibitem{Corcella:1999gs}
 G.~Corcella and M.H.~Seymour,
 hep-ph/9908388.
\bibitem{Abbott:1999yd}
  B.~Abbott {\it et al.} [D$\emptyset$ Coll.],
  hep-ex/9909020.
\bibitem{Abbott:1999th}
 B.~Abbott {\it et al.} [D$\emptyset$ Coll.],
 hep-ex/9907044.
\bibitem{Ellison:1999jx}
  J.~Ellison, [CDF and D$\emptyset$ Coll.]
  hep-ex/9910037, {\it to appear in these proceedings.}
\bibitem{Kuhlmann:1999xk}
  S.~Kuhlmann [CDF Coll.], FERMILAB-CONF-99-165-E.
\bibitem{Breitweg:1999su}
  J.~Breitweg {\it et al.} [ZEUS Coll.],
  hep-ex/9910045.
\bibitem{Acciarri:1999md}
  M.~Acciarri {\it et al.} [L3 Coll.],
  hep-ex/9909005.
\bibitem{Mueller:1986zp}
  A.H.~Mueller and P.~Nason,
  Nucl.\ Phys.\ {\bf B266}, 265 (1986).
  M.L.~Mangano and P.~Nason,
  Phys.\ Lett.\ {\bf B285}, 160 (1992).
  M.H.~Seymour,
  Nucl.\ Phys.\ {\bf B436}, 163 (1995).
\bibitem{Miller:1998ig}
  D.J.~Miller and M.H.~Seymour,
  Phys.\ Lett.\ {\bf B435}, 213 (1998)
  hep-ph/9805414.
\bibitem{LEPHF}
  The LEP/SLD Heavy Flavour Working Group, D. Abbaneo et al., LEPHF/99-01.
\bibitem{Abbiendi:1999sx}
  G.~Abbiendi {\it et al.} [OPAL Coll.],
  hep-ex/9908001.
\bibitem{Abreu:1999qh}
  P.~Abreu {\it et al.} [DELPHI Coll.],
  CERN-EP-99-081.
\bibitem{Abe:1999qg}
  K.~Abe {\it et al.}
  [SLD Coll.],
  hep-ex/9908028.
\bibitem{Adloff:1998vb}
  C.~Adloff {\it et al.} [H1 Coll.],
  Nucl.\ Phys.\ {\bf B545}, 21 (1999)
  hep-ex/9812023.
\bibitem{Breitweg:1998yt}
  J.~Breitweg {\it et al.} [ZEUS Coll.],
  Eur.\ Phys.\ J.\ {\bf C6}, 67 (1999)
  hep-ex/9807008.
\bibitem{Breitweg:1999ad}
  J.~Breitweg {\it et al.} [ZEUS Coll.],
  hep-ex/9908012.
\bibitem{Frixione:1995dv}
  S.~Frixione, M.L.~Mangano, P.~Nason and G.~Ridolfi,
  Phys.\ Lett.\ {\bf B348}, 633 (1995)
  hep-ph/9412348.
\bibitem{Frixione:1995qc}
  S.~Frixione, P.~Nason and G.~Ridolfi,
  Nucl.\ Phys.\ {\bf B454}, 3 (1995)
  hep-ph/9506226.
\bibitem{Cacciari:1997du}
  M.~Cacciari and M.~Greco,
  Phys.\ Rev.\ {\bf D55}, 7134 (1997)
  hep-ph/9702389.
\bibitem{Binnewies:1998xq}
  J.~Binnewies, B.A.~Kniehl and G.~Kramer,
  Phys.\ Rev.\ {\bf D58}, 014014 (1998)
  hep-ph/9712482.
\bibitem{Olness:1997yc}
  F.I.~Olness, R.J.~Scalise and W.~Tung,
  Phys.\ Rev.\ {\bf D59}, 014506 (1999)
  hep-ph/9712494.
\bibitem{Cacciari:1998it}
  M.~Cacciari, M.~Greco and P.~Nason,
  JHEP {\bf 05}, 007 (1998)
  hep-ph/9803400.
\bibitem{Frixione:1999xz}
  S.~Frixione,
  hep-ph/9905545.
\bibitem{Adloff:1999nr}
  C.~Adloff {\it et al.} [H1 Coll.],
  hep-ex/9909029.
\bibitem{Hayes:1999ie}
  M.~Hayes [H1 Coll.],
  hep-ex/9905033.
\bibitem{Nason:1988xz}
  P.~Nason, S.~Dawson and R.K.~Ellis,
  Nucl.\ Phys.\ {\bf B303}, 607 (1988).
  W.~Beenakker, H.~Kuijf, W.L.~van Neerven and J.~Smith,
  Phys.\ Rev.\ {\bf D40}, 54 (1989).
\bibitem{Nason:1989zy}
  P.~Nason, S.~Dawson and R.K.~Ellis,
  Nucl.\ Phys.\ {\bf B327}, 49 (1989).
  W.~Beenakker, W.L.~van Neerven, R.~Meng, G.A.~Schuler and J.~Smith,
  Nucl.\ Phys.\ {\bf B351}, 507 (1991).
\bibitem{Mangano:1992jk}
  M.L.~Mangano, P.~Nason and G.~Ridolfi,
  Nucl.\ Phys.\ {\bf B373}, 295 (1992).
\bibitem{Frixione:1997ma}
  S.~Frixione, M.L.~Mangano, P.~Nason and G.~Ridolfi,
  hep-ph/9702287;
  Nucl.\ Phys.\ {\bf B431}, 453 (1994).
\bibitem{Abbott:1999se}
  B.~Abbott {\it et al.}
  [D$\emptyset$ Coll.],
  hep-ex/9905024.
\bibitem{Abe:1998ac}
  F.~Abe {\it et al.}
  [CDF Coll.],
  FERMILAB-PUB-98-392-E.
\bibitem{Abbott:1999wu}
  B.~Abbott {\it et al.}
  [D$\emptyset$ Coll.],
   hep-ex/9907029.
\bibitem{Mangano:1997ri}
  M.L.~Mangano,
  hep-ph/9711337.
\bibitem{Mele:1991cw}
  B.~Mele and P.~Nason,
  Nucl.\ Phys.\ {\bf B361}, 626 (1991).
\bibitem{Nason:1999zj}
  P.~Nason and C.~Oleari,
  hep-ph/9903541.
\bibitem{Abe:1999fi}
  K.~Abe {\it et al.}
  [SLD Coll.],
  hep-ex/9908032.
\bibitem{Catani:1996yz}
  S.~Catani, M.L.~Mangano, P.~Nason and L.~Trentadue,
  Nucl. Phys. {\bf B478}, 273 (1996)
  hep-ph/9604351;
  Phys. Lett. {\bf B378}, 329 (1996)
  hep-ph/9602208.
\bibitem{Abbott:1998nn}
  B.~Abbott {\it et al.}
  [D$\emptyset$ Coll.],
  Phys. Rev. {\bf D60}, 012001 (1999)
  hep-ex/9808034.
\bibitem{Berger:1998gz}
  E.L.~Berger and H.~Contopanagos,
  Phys. Rev. {\bf D57}, 253 (1998)
  hep-ph/9706206.
\bibitem{Kidonakis:1998qe}
  N.~Kidonakis,
  hep-ph/9904507.
\bibitem{Koehn:1999tz}
  P.~Koehn [CDF Coll.], FERMILAB-CONF-99-306-E, Jul 1999, {\it
  to  appear in these Proceedings.}
\bibitem{Frixione:1995fj}
  S.~Frixione, M.L.~Mangano, P.~Nason and G.~Ridolfi,
  Phys. Lett. {\bf B351}, 555 (1995)
  hep-ph/9503213.
\bibitem{Kilgore:1997sq}
  W.B.~Kilgore and W.T.~Giele,
  Phys.\ Rev.\ {\bf D55}, 7183 (1997)
  hep-ph/9610433.
\end{thebibliography}
\end{document}